\documentclass[aps,
amsmath,
floats,
floatfix,
12pt,
nofootinbib,
tightenlines,
showpacs,
showlabels,
superscriptaddress]{revtex4-1}
\usepackage{mathrsfs}
\usepackage{fancyvrb}
\usepackage{amsmath}
\usepackage{amssymb}
\usepackage{graphicx}
\usepackage{accents}
\usepackage{algorithm}
\usepackage{algpseudocode}
\usepackage{setspace} 
\newcommand{\sbt}{{}_{\vcenter{\hbox{\large \bfseries .}}}}
\renewcommand{\dot}[1]{{\stackrel{\smash[b]\sbt}{#1}}{}}

\newcommand{\uright}{W}
\newcommand{\uleft}{X}
\newcommand{\scri}{\mathscr{I}^+}
\newcommand{\unm}{Mathematics and Statistics,
                  University of New Mexico,
                  Albuquerque, NM 87131, USA.}
\newcommand{\umd}{Department of Physics, 
                  Maryland Center for Fundamental Physics and
                  Joint Space Science Institute,
                  University of Maryland, 
                  College Park, MD 20742, USA.}
\begin{document}

\title{Fast evaluation of asymptotic waveforms\\ 
       from gravitational perturbations}

\author{Alex G.~Benedict${}^{1}$, 
        Scott E.~Field${}^{\,2}$, 
        Stephen R.~Lau}
\affiliation{\unm\\[3mm]
             ${}^2$\umd}
\date{\today}

\begin{abstract}

In the context of blackhole perturbation theory, we describe
both exact evaluation of an asymptotic waveform from 
a time series recorded at a finite radial location and its
numerical approximation.
From the user's standpoint our technique is easy to 
implement, affords high accuracy, and works for both axial 
(Regge-Wheeler) and polar (Zerilli) sectors. Our focus is on the 
ease of implementation with publicly available numerical tables, 
either as part of an existing evolution code or a post-processing
step. Nevertheless, we also present a thorough 
theoretical discussion of asymptotic waveform evaluation and 
radiation boundary conditions, which need not be understood by a 
user of our methods. In particular, we identify (both in the time 
and frequency domains) analytical asymptotic waveform evaluation 
kernels, and describe their approximation by techniques developed 
by Alpert, Greengard, and Hagstrom. This paper also presents 
new results on the evaluation of far-field signals for the 
ordinary (acoustic) wave equation. We apply our method 
to study late-time decay tails at null-infinity, ``teleportation" 
of a signal between two finite radial values, and luminosities 
from extreme-mass-ratio binaries. Through numerical simulations 
with the outer boundary as close in as $r=30M$, we compute 
asymptotic waveforms with late-time $t^{-4}$ decay
($\ell = 2$ perturbations), and also luminosities 
from circular and eccentric particle-orbits 
that respectively match frequency domain results to relative 
errors of better than $10^{-12}$ and $10^{-9}$.
Furthermore, we find that asymptotic waveforms are especially
prone to contamination by spurious junk radiation.
\end{abstract}
\maketitle

\section{Introduction} \label{sec:intro}
Asymptotic-waveform evaluation (AWE) is a long standing challenge in the 
computation of waves. Whether for acoustic, electromagnetic, 
or gravitational waves the goal is to identify the far-field 
or asymptotic signal radiated to future null infinity $\scri$ 
using only knowledge of the 
solution on a truncated, spatially 
finite, computational domain. For the unit-speed ordinary
wave equation, the far-field signal is $f(u,\theta,\phi) = 
\lim_{r\rightarrow\infty}r\psi(u,r,\theta,\phi)$, where
$\psi(u,r,\theta,\phi)$ is a solution to the wave equation
written with respect to retarded-time $u=t-r$ and spherical 
polar coordinates $(r,\theta,\phi)$. Continuing with this 
example, we view computation of 
$r_\infty \psi(u,r_\infty,\theta,\phi)$ as AWE, so long as
the {\em evaluation radius} $r_\infty$ can be taken arbitrarily 
large, even if ultimately finite. Indeed, for the ordinary 
wave equation the asymptotic signal and signal at 
$r_\infty = 10^{15}$ are identical to about double precision machine 
epsilon (see Appendix~\ref{sec:ErrorEst} for the error estimate).
The asymptotic signal we compute corresponds to
observation on the timelike hyper-cylinder $r = r_\infty$.
In the perturbative gravitational setting considered here, with 
$r_\infty = 2M \left(1 \times 10^{15}\right)$ in terms of
the Schwarzschild mass $M$, such observations take place 
in the {\em astrophysical zone}~\cite{Leaver,Barack1999,Purrer:2004nq}.
An observation in the astrophysical zone is better approximated
as taking place at $\scri$ rather than future
timelike infinity $i^+$ \cite{Zenginoglu:2008wc,Leaver,Purrer:2004nq,Barack1999}.
For this reason, and for clarity of exposition, throughout this 
paper we refer to such observations as taking place at $\scri$.

Identifying the gravitational wave signal at $\scri$ is of 
both theoretical and practical importance. Theoretically, in 
the context of asymptotically flat spacetimes Sachs \cite{Sachs:1962} 
identified the asymptotic metric factors corresponding to 
$f(u,\theta,\phi)$ for gravitation, and exploited this
identification in his discussion of the radiative degrees 
of freedom for general relativity. Using Geroch's 
calculation framework \cite{Geroch}, Ashtekar and Streubel 
expanded on the Sachs approach in their fundamental analysis 
\cite{AshtekarStreubel} of the symplectic structure of 
radiative modes and gravitational flux in general relativity. 
Several works \cite{DrayStreubel,Dray,Shaw} 
then investigated the general charge integral (where the 
integration is over a two-surface ``cut" of $\scri$) corresponding 
to the Ashtekar-Streubel flux. The practical importance of AWE stems 
from the upcoming generation of advanced-sensitivity ground-based 
gravitational wave interferometer detectors 
(i.e., advanced LIGO, advanced Virgo, and KAGRA)
\cite{KAGRA_web,LIGO_web,VIRGO_web,GEO_web} 
and anticipated space-based detectors 
like LISA \cite{LISA_web,ESA_web}. These 
instruments are well-modeled as idealized observers located 
at $\scri$. 

While the problem of AWE for 
gravitation shares difficulties with its counterparts for 
the ordinary wave and Maxwell equations, for example the slow 
fall-off of the waves (in our case metric perturbations) in 
powers of $1/r$, the gravitational problem is further 
complicated by the backscattering of waves, coordinate 
(gauge) issues, and non-linearities.
For a perturbed Schwarzschild blackhole, covariant and gauge 
invariant approaches exist for the construction of ``master
functions" from the spacetime metric perturbations (see, for 
example, \cite{Sarbach:2001qq,Martel_CovariantPert,SopuertaLaguna}).
In the asymptotic limit these master functions specify the
gravitational waveform. Here we consider the wave equations
which directly govern these master functions.
In this perturbative setting, 
AWE is precisely the technique (perhaps extrapolation, 
for example) used to compute the master functions at 
arbitrarily large distances from the central blackhole. 
This paper introduces a new technique based on signal 
{\em teleportation} between two finite radial values.

A straightforward and longstanding approach to AWE in both 
full general relativity \cite{Pollney:2009yz,Boyle2009} 
and perturbative settings \cite{Sundararajan:2007jg} has been 
to record relevant field quantities at a variety of radii, 
perform a numerical fit, and then extrapolate to larger radii. 
However, the accuracy of this method ultimately relies on an 
{\em Ansatz} for the expected fall-off of the field with 
larger $r$, as well as recording field values at multiple and 
preferably large values of $r$ \cite{Pollney:2009ut}. 
An alternative approach, known as Cauchy characteristic 
extraction, is to record geometrical data from a Cauchy evolution 
on a world-tube, which is later used as interior boundary data 
for a second characteristic evolution whose coordinates have been 
compactified to formally include $\scri$ within the numerical 
grid~\cite{Bishop,Reisswig:2009us,Babiuc:2010ze,Winicour_LRR}.
When the background coordinates are fixed, $\scri$ can 
be directly included within a Cauchy evolution by a
geometric prescription using hyperboloidal methods
\cite{Zenginoglu:2008wc,Zenginoglu:2007jw,Zenginoglu:2009hd,Zenginoglu:2009ey,
Zenginoglu:2010cq,Zenginoglu:2010zm,Bernuzzi:2011aj,Zenginoglu:2011zz}.
Another approach due to Abrahams and Evans shows 
how one may {\em exactly} evaluate asymptotic waveforms 
from gravitational multipoles for general relativity 
linearized about flat spacetime \cite{AE1,AE2}.
This paper presents a new analytical and 
numerical method to evaluate asymptotic 
gravitational waveforms from perturbations of a 
non-spinning (Schwarzschild) 
blackhole. It also presents new results on the asymptotic signal 
evaluation problem for the acoustic (i.e.~ordinary) wave equation. 
Our approach is most similar to that of Abrahams and Evans. 
Essentially, we reformulate their approach in a way which 
subsequently generalizes to blackhole 
perturbations. However, while our approach generalizes 
the Abrahams-Evans one to a curved background spacetime, 
we do not match their careful discussion of gauge issues.

For the Einstein equations linearized about Minkowski spacetime in
the Lorenz gauge the trace-reversed metric perturbation obeys the 
flatspace (ordinary d'Alembertian) tensor wave equation. Therefore, 
these perturbations are akin to solutions of either the ordinary 
wave equation or the Maxwell equations; solutions characterized by
the sharp Huygen's principle and, therefore, which possess secondary 
lacunae \cite{PetropavlovskyTsynkov}: given trivial initial 
data and an inhomogeneous source which is bounded in space and time, 
the solution vanishes on the intersection of all forward light cones 
whose vertices sweep over the support of the source. The secondary 
lacunae is a region of spacetime which is ``dark" because all 
waves have already passed. Similar statements hold for the
homogeneous case with non-trivial initial data of compact support.
Actually, for the Maxwell case with certain sources, the solution may 
have a quasi-lacunae featuring a late-time static electric field 
\cite{PetropavlovskyTsynkov}.

Wave propagation on a curved spacetime 
is more complicated due to the backscattering of waves off of curvature. 
Even within the relatively simple setting of perturbations of 
Schwarzschild blackholes, backscattering effects are present and 
the resulting late-time ``tails" \cite{Price1972,DSS2011}
have been extensively studied both theoretically and numerically. 
Backscattering confounds our intuitive sense of ``outgoing" and 
``ingoing"; one might reasonably take the viewpoint that a partially 
backscattered wave has both outgoing and ingoing pieces. Nevertheless, 
for the {\em linear} master equations which describe perturbations of 
Schwarzschild blackholes, there is an unambiguous notion of ``outgoing", 
provided initial data of compact support. Away from the support of the 
initial data, Laplace transformation of a master equation 
\eqref{eq:RWZeqn} yields a {\em homogeneous} second-order ODE, which 
therefore has two linearly independent solutions,
$\widehat{\Psi}^{(1)}(s,r)$ and $\widehat{\Psi}^{(-1)}(s,r)$,
where $s$ is Laplace frequency. Here we have suppressed 
harmonic indices $(\ell,m)$ and assumed that the area radius $r$
is the independent spatial variable. We may assume 
that $\widehat{\Psi}^{(\pm 1)}(s,r) \sim \exp(\mp sr_*)$ as 
$r, r_* \rightarrow \infty$, where $r_*$ is the Regge-Wheeler 
tortoise coordinate defined below. 
At a radial location beyond the support 
of the initial data, the frequency-domain solution has the
form $\alpha(s)\widehat{\Psi}^{(+1)}(s,r)$, where the details of
the initial data are buried in the coefficient $\alpha(s)$.
Physically, this notion of ``outgoing" would perhaps be better
characterized as ``asymptotically outgoing". Nevertheless,
provided the solution has this form, we can derive {\em at a finite 
radius} both time-domain boundary conditions 
\cite{Lau1,Lau2,Lau3} and an AWE procedure. The strategy in
both cases is to write down the exact conditions/procedure in the
frequency domain, and then accurately approximate this exact 
relationship in a fashion that allows for simple inversion under
the inverse Laplace transform. For the case of boundary conditions,
one approximates the exact Dirichlet-to-Neumann map as a rational
function (in fact a sum of simple poles) along the axis of
imaginary Laplace frequency (the inversion contour). The exact 
time-domain boundary condition is a history-dependent convolution,
which maybe approximated to machine precision as a
convolution involving a kernel given by a small sum of 
exponentials. As we show, this type of kernel effectively 
localizes the history dependence.

Our reformulation of the Abrahams-Evans procedure
(and its generalization to curved spacetimes) features a similar 
history-dependent convolution involving a sum-of-exponentials 
time-domain kernel. Section \ref{sec:Theory} demonstrates that, 
in the (Laplace) frequency domain, an AWE 
kernel is exactly expressible as an ``integral over boundary kernels", 
thereby allowing us to leverage existing codes and knowledge 
for generating and approximating boundary kernels. While 
the construction of AWE kernels is 
computationally intensive, this is an {\em offline} cost. Once the 
kernel has been calculated, efficient and accurate AWE
can be implemented within an existing code in a non-intrusive manner.
Furthermore, AWE can be effected as a post-processing 
step on existing data recorded at a fixed radial location. 
Kernels used in this paper, as well as others, will be 
available at \cite{Kernel_web1}.

This paper is organized as follows. Section \ref{sec:HowTo}
provides a self-contained guide on using 
boundary and AWE kernels in either 
existing codes or data post-processing. Towards this end, 
Section~\ref{subsec:QNRTails} considers the numerical evolution 
of late time tails from an approximate asymptotic signal which we find to
decay at the rate predicted for $\ell = 2$ perturbations
at $\scri$.
Section \ref{sec:Theory} presents the theoretical underpinnings
of both radiation boundary conditions and waveform teleportation, 
considering both wave propagation on flat (Minkowski) and 
Schwarzschild spacetime.
For $\ell = 2$, $3$, $64$ perturbations, Section~\ref{subsec:PulseT}
considers accurate signal teleportation to a finite (near-field)
radial value. In Section~\ref{subsec:luminosities} we apply our 
method to compute gravitational waveforms and luminosities from 
extreme mass ratio binary systems, finding excellent 
agreement with frequency domain computations. In these studies 
we have observed that spurious junk radiation is 
problematic for accurate $\scri$ computations. Finally, we
conclude in Section \ref{sec:conclude} by discussing open issues,
both theoretical and practical.

\section{Implementation how-to guide} \label{sec:HowTo}

Our aim in this section is not to give a
derivation of our AWE method. Rather, adopting the 
simplest possible evolution scheme and coordinates, we focus on 
how AWE is implemented. By presenting an 
implementation of our AWE method for a simple scheme, we hope to convey 
the key points 
to the reader, who will then grasp how to implement the method 
within their own evolution scheme. Since our implementation of AWE 
relies on certain radiation boundary conditions (RBC), 
themselves essential when working on a spatially finite domain, 
we first describe how to implement these within our simple scheme. 
Here we do not discuss our RBC and AWE methods for different 
background coordinate systems (i.e.~Kerr-Schild or hyperboloidal 
foliations), but return to this issue in Appendix~\ref{sec:OtherFol}.

Multipole gravitational perturbations of a Schwarzschild blackhole are 
described by the Regge-Wheeler (axial) and Zerilli (polar) 
formalisms \cite{RW,ZER}. In geometric units the corresponding ``master" 
wave equations have the form
\begin{equation}\label{eq:RWZeqn}
       \frac{\partial^2\Psi}{\partial t^2}
                  -\frac{\partial^2\Psi}{\partial r_*^2} +
                   V^\mathrm{RW,Z}(r)\Psi = S,
\end{equation}
where $S(t,r)$ is a possible source and in terms of 
the blackhole mass $M$ the Regge-Wheeler tortoise coordinate 
is $r_* = r + 2M\log(\frac{1}{2}M^{-1}r - 1)$, which we also 
denote by $x$. Until Section \ref{subsec:luminosities}, 
we alway choose $S = 0$ for the source.
Expressions for the Zerilli $V^\mathrm{Z}(r)$ 
and Regge-Wheeler $V^\mathrm{RW}(r)$ potentials are given 
in, for example, Eqs.~(2) and (3) of Ref.~\cite{FHL}. 
Both potentials depend on the orbital angular index 
$\ell$. We have suppressed this index on $V(r)$, as well as the 
orbital and azimuthal indices $(\ell,m)$ on the mode $\Psi$
and source $S$.

\subsection{Simple evolution algorithm with RBC}
\label{subsec:simpleRBC}
Most numerical schemes for evolution of \eqref{eq:RWZeqn} employ
some form of ``auxiliary variables", for example the variables
$\Psi,\Pi \equiv -\partial_t\Psi,\Phi \equiv \partial_x \Psi$.
However, for transparency here we use ``characteristic 
variables":
\begin{align}\label{eq:charvars}
\Psi, \qquad \uright = -\Pi - \Phi, \qquad \uleft = -\Pi +\Phi ,
\end{align}
for which the evolution equations are
\begin{align} \label{eq:UPsiSystem} 
\partial_t \Psi = \frac{1}{2}(\uright + \uleft),\quad
\partial_t \uright  = -\partial_x \uright - V(r(x))\Psi,\quad
\partial_t \uleft  = \partial_x \uleft - V(r(x))\Psi.
\end{align}
These equations show that $\uright$ propagates left-to-right ($\nearrow$)
and $\uleft$ propagates right-to-left ($\nwarrow$). While
from the theoretical standpoint we prefer to view fields, 
for example $\Psi(t,r)$, as depending spatially on $r$, from a 
computationally standpoint we discretize Eqs.~\eqref{eq:UPsiSystem} 
in $x$. Therefore, suppose the computational domain is the interval 
$[a,b]$ in tortoise coordinate $x$, corresponding to the interval 
$[r_a, r_b]$ in Schwarzschild radius $r$. Boundary conditions must 
be specified for $\uright$ at $x=a$ and for $\uleft$ at $x=b$. 
We discretize \eqref{eq:UPsiSystem} using first-order 
upwind stencils in space, and the forward Euler method in time. 
Respectively, let $\{t_n\}$ and $\{x_k\}_{k=0}^K$ denote uniformly 
spaced temporal and spatial grid points 
such that $\Delta x = x_{k+1}-x_k$. Moreover, let $V_k \equiv
V(r(x_k))$ and $\Psi^n_k \simeq \Psi(t_n,r(x_k))$, with the same 
notation for the other fields. Then the scheme for updating the 
fields at all $x_k$ from time $t_n$ to time $t_{n+1} = 
t_n + \Delta t$ is given by Algorithm~\ref{alg:FD_scheme}.
{\scriptsize
\begin{algorithm}[H]
\caption{Finite difference scheme for Eqs.~\eqref{eq:UPsiSystem}}
\label{alg:FD_scheme}
\begin{algorithmic}[1]
\State $\Psi^{n+1}_0    = \Psi^n_0
                        + (\Delta t/2) 
                         \big[\uright^n_0 + \uleft^n_0\big]$
\State $\uright^{n+1}_0 = \uright_a^{n+1}$ 
                         \Comment{{\bf Boundary condition at} $a$}
\State $\uleft^{n+1}_0  = \uleft^n_0  
                       + (\Delta t/\Delta x)
                         \big(\uleft^n_1-\uleft^n_0\big) 
                       - \Delta t V_0\Psi^n_0$
\For{$k = 1 \text{ \bf{to} } K-1$}
\State \hspace{5mm} $\Psi^{n+1}_k     = \Psi^n_k 
                                     + (\Delta t/2) 
                                     \big[\uright^n_k + \uleft^n_k\big]$
\State \hspace{5mm} $\uright^{n+1}_k  = \uright^n_k  
                                     - (\Delta t/\Delta x)
                                       \big(\uright^n_k-\uright^n_{k-1}\big) 
                                     - \Delta t V_k \Psi^n_k$
  \State \hspace{5mm} $\uleft^{n+1}_k = \uleft^n_k  
                                     + (\Delta t/\Delta x)
                                       \big(\uleft^n_{k+1}-\uleft^n_k\big) 
                                     - \Delta t V_k\Psi^n_k$
\EndFor
\State $\Psi^{n+1}_K     = \Psi^n_K 
                         + (\Delta t/2) 
                          \big(\uright^n_K + \uleft^n_K\big)$
\State $\uright^{n+1}_K  = \uright^n_K 
                        - (\Delta t/\Delta x)
                          \big(\uright^n_K-\uright^n_{K-1}\big) 
                        - \Delta t V_K \Psi^n_K$
\State $\uleft^{n+1}_K   = \uleft_b^{n+1}$  
                          \Comment{{\bf Boundary condition at} $b$}
\end{algorithmic} 
\end{algorithm}
}

To complete the scheme, we must specify both 
$\uright_a^n \simeq W(t_n,r_a)$ and $\uleft_b^n \simeq X(t_n,r_b)$
as functions of (discrete) time. 
So long as $a \ll 0$, the inner boundary value of the potential 
$V(r)$ near $r = r_a$ is zero to machine precision, and the 
Sommerfeld boundary condition $\uright_a^n = 0, \forall n$ 
is highly accurate. The RBC at
$x=b$ is determined by a Laplace convolution,
\begin{align}\label{eq:LaplaceConvApprox}
\uleft(t,r_b) = \frac{f(r_b)}{r_b}
\int_0^t \Xi(t-t',r_b)
\Psi(t',r_b)dt',
\end{align}
where $f(r) = 1-2M/r$ and the boundary time-domain {\em kernel} is
\begin{equation}\label{eq:compressedBHkern}
\Xi(t,r_b) = \sum_{q=1}^d \frac{\gamma_q(\rho_b)}{2M} \Xi_q(t,r_b), \qquad
\Xi_q(t,r_b) = \exp\left(\frac{\beta_q(\rho_b)t}{2M}\right).
\end{equation}
The parameters $\{(\gamma_q(\rho_b),\beta_q(\rho_b))\}_{q=1}^d$ 
depend on the rescaled boundary radius $\rho_b = (2M)^{-1}r_b$
(as well as the orbital index $\ell$ which is suppressed) and 
they are listed in numerical tables, such as those given in 
the Appendix \ref{App:Tables}.\footnote{These 
parameters can be redefined through division by $2M$, 
thereby removing $2M$ factors in the formulas which follow. Indeed, 
such a redefinition would be made in an actual code. Nevertheless, we 
retain the original $\{(\gamma_q(\rho_b),\beta_q(\rho_b))\}_{q=1}^d$
parameters with $2M$ factors, in order to ensure that the 
parameters here correspond precisely to those listed in our
tables.} Some of the parameters $\{(\gamma_q,\beta_q)\}_{q=1}^d$ 
are complex, but the kernel $\Xi(t,r_b)$ is real. We stress that, 
insofar as implementation of the convolution 
\eqref{eq:LaplaceConvApprox} is concerned, the origin of 
these numbers is unimportant. Defining, for example, 
$\Psi_b(t) = \Psi(t,r_b)$, we write \eqref{eq:LaplaceConvApprox} 
as $X_b(t) = r_b^{-1}f(r_b)(\Xi(\cdot,r_b) * \Psi_b)(t)$. With 
a similar notation, the constituent 
convolution $(\Xi_q * \Psi_b) \equiv 
(\Xi_q(\cdot,r_b) * \Psi_b)(t)$ obeys an ODE 
{\em at the boundary},
\begin{equation}\label{eq:boundaryODE}
\frac{d}{dt}
(\Xi_q * \Psi_b) = \big[(2M)^{-1}\beta_q (\Xi_q * \Psi_b) + \Psi_b\big].
\end{equation}
These $d$ ODE can be integrated along side the system 
\eqref{eq:UPsiSystem}. Indeed, we define $(\Xi_q * \Psi_b)^n \simeq
(\Xi_q * \Psi_b)(t_n)$ and complete our scheme as follows.
{\scriptsize
\begin{algorithm}[H]
\caption{Finite difference scheme for Eqs.~\eqref{eq:UPsiSystem} with RBC}
\label{alg:RBCs}
\begin{algorithmic}[1]
\For{$q = 1 \text{ \bf{to} } d$} \Comment{{\bf First update constituent convolutions}}
\State \hspace{5mm} $(\Xi_q * \Psi_b)^{n+1} = (\Xi_q * \Psi_b)^n 
                                         + \Delta t \big[(2M)^{-1}\beta_q (\Xi_q * \Psi_b)^n 
                                         + \Psi^n_K\big]$
\EndFor
\State $\uleft_b^{n+1} = (2Mr_b)^{-1}f(r_b)\sum_{q=1}^d 
                         \gamma_q (\Xi_q * \Psi_b)^{n+1}$
\State $\uright_a^{n+1} = 0$
\vskip 10pt
\State Run Algorithm~\ref{alg:FD_scheme} with 
       $\uright_a^{n+1}$ and 
       $\uleft_b^{n+1}$ given by above values.
\end{algorithmic}
\end{algorithm}
}

\subsection{Evaluation of the asymptotic waveform} 
\label{subsec:simpleEXT}

Our approach to AWE is similar to the described implementation 
of radiation boundary conditions. We introduce new parameters 
$\{(\gamma^E_q(\rho_b,\rho_\infty),
    \beta^E_q(\rho_b,\rho_\infty))
\}_{q=1}^{d^E}$ and a new kernel 
\begin{equation}\label{eq:XiEkernel}
\Xi^E(t,r_b,r_\infty) = 
\sum_{q=1}^{d^E}\frac{\gamma^E_q(\rho_b,\rho_\infty)}{2M}
\Xi^E_q(t,\rho_b,\rho_\infty),\qquad 
\Xi^E_q(t,\rho_b,\rho_\infty) = 
\exp\left(\frac{\beta^E_q(\rho_b,\rho_\infty)t}{2M}\right).
\end{equation}
The parameters $\{(\gamma^E_q(\rho_b,\rho_\infty), 
\beta^E_q(\rho_b,\rho_\infty)\}_{q=1}^{d^E}$
also depend on $\ell$, but this dependence has been suppressed.
The $E$ here stands for ``evaluation"
and differentiates this kernel 
from the RBC one. This evaluation kernel enacts {\em teleportation}
(the term is defined in Section~\ref{sec:Theory}) of the waveform 
from the boundary radius $r_b = 2M\rho_b$ to the evaluation
radius $r_\infty = 2M\rho_\infty \gg r_b$ (from $b$ to $x_\infty$ 
in tortoise coordinate). As discussed below, the 
choice $r_\infty = \infty$ is formally 
possible (see Sec.~\ref{sec:conclude}); however, in this
paper $r_\infty$ is an {\em arbitrarily large}, albeit finite, 
radius. To ensure the signals recovered at $r_\infty$ and $\scri$ 
are identical to about machine precision, 
we choose $r_\infty = 2M(1\times 10^{15})$ for double precision 
simulations, and would choose $r_\infty = 2M(1\times 10^{30})$
for quadruple precision simulations. We then 
approximate the asymptotic waveform as
\begin{equation}\label{eq:XiEkernelConv}
\Psi_\infty(t) := \Psi(t+(x_\infty-b),r_\infty) \simeq 
\int_0^t \Xi^E(t-t',r_b,r_\infty)\Psi(t',r_b)dt' + \Psi(t,r_b).
\end{equation}
The offset by $\Psi(t,r_b)$ in this formula stems from a 
technicality explained in Section~\ref{Sec:flatspace}.
As before, this formula can be implemented through integration 
of ODE at the boundary, only now these ODE are not coupled to 
the numerical evolution. With $(\Xi_q^E * \Psi_b)^n
\simeq (\Xi_q^E(\cdot,r_b) * \Psi_b)(t_n)$ and 
$\Psi_\infty^{n} \simeq \Psi_\infty(t_n)$, the algorithm 
using forward Euler is as follows.
{\scriptsize
\begin{algorithm}[H]
\caption{Evaluation of asymptotic waveform, placed after field update}
\label{alg:WEs}
\begin{algorithmic}[1]
\For{$q = 1 \text{ \bf{to} } d^E$}
\State \hspace{5mm} $(\Xi_q^E * \Psi_b)^{n+1} 
                       = (\Xi_q^E * \Psi_b)^n 
                       + \Delta t \big[ 
                         (2M)^{-1}\beta_q^E (\Xi_q^E * \Psi_b)^n 
                       + \Psi^n_K\big]$
\EndFor
\State $\Psi_\infty^{n+1} = (2M)^{-1}\sum_{q=1}^d \gamma_q^E 
                            (\Xi_q^E * \Psi_b)^{n+1} 
                          + \Psi^{n+1}_K$
\end{algorithmic}
\end{algorithm}
}
\noindent
The following condition would yield particularly efficient AWE:
\begin{equation}\label{eq:WEandRBCpoles}
d^E = d,\quad
\beta^E_q(\rho_b,\rho_\infty) = \beta_q(\rho_b),\quad \forall q,
\text{ (preferred, but perhaps {\em not} possible).}
\end{equation} 
Indeed, integration of the same ODE \eqref{eq:boundaryODE} at the 
boundary would then determine both the RBC and AWE. In this case, 
steps 1 through 3 of Algorithm \ref{alg:WEs} have already been 
carried out in Algorithm \ref{alg:RBCs}. However, the assumption 
in \eqref{eq:WEandRBCpoles} may not always be possible, and even
when possible appears to yield less accurate teleportation kernels. We have 
constructed $\ell = 2$ AWE kernels which satisfy \eqref{eq:WEandRBCpoles} 
(an example is given in the Appendix \ref{App:Tables}); however, 
relative to our best kernels, they indeed yield less accuracy.
Moreover, we have been unable to achieve \eqref{eq:WEandRBCpoles}
for $\ell=64$ teleportation kernels. Therefore, in this paper
we will {\em not} assume \eqref{eq:WEandRBCpoles}.
\vskip 10pt

\noindent
{\bf Remark.} If $\Psi$ is itself complex (as is the case in 
applications), then round-off issues will lead to mixing of the 
real and imaginary parts in the simple algorithms above.
In this case, we advocate splitting the complex exponentials which 
make up $\Xi$ and $\Xi^E$ into manifestly real expressions 
involving sine and cosine terms. Such splitting amounts to extra 
bookkeeping, but hardly complicates the above treatment.
\begin{figure}[t]
\includegraphics[clip=true,width=14cm]{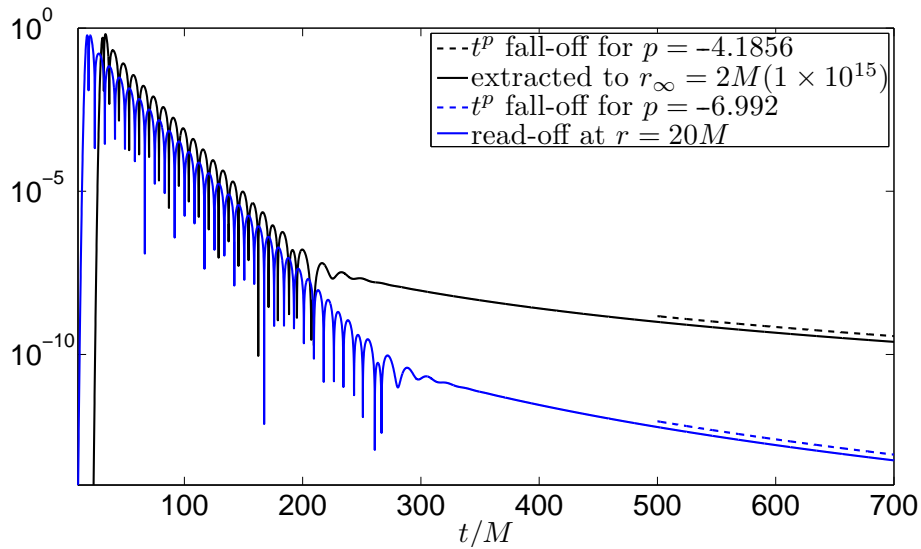}
\caption{{\sc Quasinormal ringing and decay tails.}
Each dashed curve corresponds to power law decay, with the
indicated rate determined by a least squares fit of 
the field over the time window $[500,700]M$. 
The time shift for the asymptotic waveform has not been included.
\label{fig:decaytails}}
\end{figure}

\subsection{Numerical experiment: quasinormal ringing and decay tails}
\label{subsec:QNRTails} 
Section \ref{sec:Experiments} documents the results of
several numerical experiments which validate and test 
our methods. This subsection also describes a numerical 
experiment, although here with the goal of providing 
further assistance toward implementation. An interested 
reader might first repeat the experiment described below.

We consider the $\ell = 2$ Regge-Wheeler equation and 
Gaussian initial data
\begin{equation}\label{eq:GaussianData}
\Psi = e^{-[2-x/(2M)]^2},\quad
\Phi = \frac{4M-x}{2M^2}e^{-[2-x/(2M)]^2},\quad
\Pi = \Phi(0,r(x)).
\end{equation}
For this experiment the computational $x$-domain is $[a,b]$, 
where $a = -200M$ and $b = 30M + 2M\log(14)$ corresponds to 
$\rho_b = r_b/(2M) = 15$. We evolve the data 
\eqref{eq:GaussianData} until time $t = 600M$, 
with a Sommerfeld boundary condition at $x=a$ and 
the convolution boundary condition \eqref{eq:LaplaceConvApprox} 
at $x=b$. Table \ref{tab:appendixRBCtable} in Appendix \ref{App:Tables}
lists the 26 pole RBC kernel $\Xi$ used for this convolution.
Instead of Algorithm~\ref{alg:FD_scheme} we use 
a multidomain Chebyshev collocation method with classical 
Runge Kutta 4 as the timestepper. 

During the simulation we record as a time series both the field 
$\Psi(t,20M)$ and $\Psi(t,r_b)$. Through the convolution 
\eqref{eq:XiEkernelConv} determined by Table 
\ref{tab:appendixEXTtableLong} in Appendix \ref{App:Tables}, 
the field $\Psi(t,r_b)$ is teleported from $r_b = 30M$
to $r_\infty = 2M(1\times 10^{15})$ providing a time series 
which approximates the asymptotic waveform $\Psi_\infty(t)$. 
In absolute value $\Psi(t,20M)$ (solid blue line) and 
$\Psi_\infty(t)$ (solid black line) are depicted in 
Fig.~\ref{fig:decaytails}. The time shift for the 
teleported waveform has not been included, and we have chosen 
to record $\Psi(t,20M)$ at $r = 20M < r_b$ only to ensure that 
the time series in the plot do not lie on top of each other 
at early times. These series exhibit the phenomena of 
quasinormal ringing and late time decay tails. Each dashed
curve in the figure corresponds to power law decay, with the
indicated rate determined by a least squares fit based on 
the numerical decay of the field over dashed curve's time 
window. The decay rates $p=-7$ and $p=-4$ are respectively the 
theoretical predictions \cite{gundlach1994late,Barack:1998bw,Zenginoglu:2009ey}
for a finite radius and $\scri$. 

\begin{figure}[t]
\includegraphics[clip=true,width=13cm,trim=0 7.25cm 0 0cm]{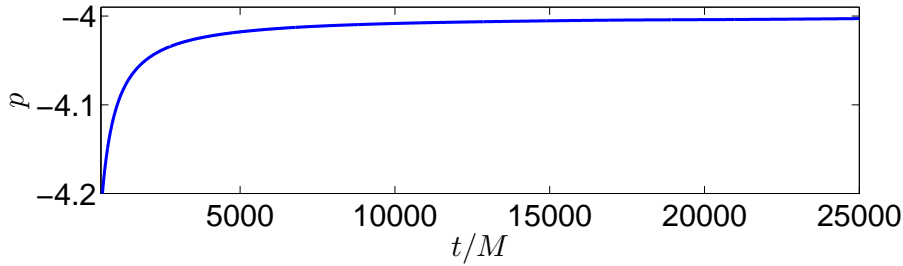}
\caption{{\sc Tail decay rate for the teleported
signal.} The rate $p$ for $\Psi_\infty(t) \propto t^p$ has been computed using
logarithmic difference quotients based on 
$\partial_{\ln t}\ln|\Psi_\infty(t)|$.
\label{fig:scritail}}
\end{figure}

Figure  \ref{fig:scritail} shows the decay rate computed from
$p = \partial_{\mathrm{ln} t} \mathrm{ln} | \Psi_\infty(t) |$
for the teleported signal with the evolution carried out to 
$t=25000M$. As a post-processing procedure, 
generation of this figure from existing data would take a few seconds.
The decay for the teleported signal asymptotically approaches $p=-4$ at 
later times. The teleported pulse corresponds to a time series recorded
along the wordline $(t_\mathrm{obs}+x_\infty-b,r_\infty)$ by an
observer safely within the astrophysical zone. Here $x_\infty-b$
is the time shift, and the astrophysical zone is defined as the region 
where $t_\mathrm{obs} \ll r_\infty$, with $t_\mathrm{obs}$ the time 
elapsed after the pulse's leading edge passes our fictitious 
observer \cite{Leaver,Purrer:2004nq,Barack1999}. Since 
an observation in the astrophysical zone is well approximated as taking 
place at $\scri$, the observed $-4$ decay rate is expected. For
very late times the decay rate should settle towards $-7$, 
although extended precision might be necessary to capture the transition.

\section{Theoretical discussion} \label{sec:Theory}

To fix ideas and motivate the new method, the next subsection 
describes AWE for flatspace multipole solutions 
of the ordinary 3+1 wave equation, formally the $M=0$ case of 
Eq.~\eqref{eq:RWZeqn}. Formulas derived in the next 
subsection motivate similar ones given for blackhole 
perturbations in subsection \ref{subsec:BH_perts}.

\subsection{Flatspace waves} \label{Sec:flatspace}
This subsection describes
(i) outgoing multipole solutions to the flatspace radial 
wave equation, (ii) the exact RBC obeyed by outgoing 
multipoles, and (iii) the relationship between this RBC 
and AWE for outgoing multipoles. Throughout this subsection, 
we often choose $\ell=2$ as a representative example, 
but similar (obvious) results hold for any multipole-order $\ell$.

\subsubsection{Structure of outgoing and ingoing flatspace
multipoles}
General outgoing ($\epsilon = 1$) and ingoing ($\epsilon = -1$)
order-$\ell$ multipole solutions of the 3+1 wave equation 
$(-\partial_t^2+\partial_x^2 + \partial_y^2 + 
\partial_z^2)\psi = 0$ have the form (see, for example,
Refs.~\cite{Burke,gundlach1994late})
\begin{equation}\label{eq:defpsil}
\psi(t,x,y,z)
= \frac{1}{r}\Psi^{(\epsilon)}_\ell(t,r) 
Y_{\ell m}(\theta,\phi),
\quad
\Psi^{(\epsilon)}_\ell(t,r) = \sum_{k=0}^\ell
\frac{\epsilon^k}{r^{k}}c_{\ell k}
f^{(\ell-k)}(t-\epsilon r),
\quad
c_{\ell k} = \frac{1}{2^k k!}
\frac{(\ell+k)!}{(\ell-k)!},
\end{equation}
where we have suppressed the azimuthal index $m$ on the
``mode" $\Psi^{(\epsilon)}_\ell(t,r)$. In 
Eq.~\eqref{eq:defpsil}
$f^{(p)}(u)$ is the $p$th derivative of an underlying
function $f(u)$ of retarded time $u = t - r$, and similarly
for $f^{(p)}(v)$, where $v = t + r$ is advanced time. In
(\ref{eq:defpsil}) we view $(x,y,z)$ as place holders for
$(r\sin\theta\cos\phi,r\sin\theta\sin\phi,r\cos\theta)$. 
For both $\epsilon = \pm 1$ cases the mode 
obeys the {\em flatspace radial wave equation}
\begin{equation}\label{eq:flatradial}
\frac{\partial^2\Psi^{(\epsilon)}_\ell}{\partial t^2}
-\frac{\partial^2\Psi^{(\epsilon)}_\ell}{\partial r^2}
+\frac{\ell(\ell + 1)}{r^2}\Psi^{(\epsilon)}_\ell = 0 ,
\end{equation}
and, specializing to the representative example, the outgoing
quadrupole is
\begin{equation}\label{eq:ell2mult_td}
\Psi^{(1)}_2(t,r) = f''(t-r)
                              + \frac{3}{r}f'(t-r)
                              + \frac{3}{r^2}f(t-r).
\end{equation}
We are interested in the $(\epsilon = 1)$ 
outgoing case, and so now write 
$\Psi_\ell(t,r)$ to mean $\Psi^{(1)}_\ell(t,r)$. 

Given a fixed radius $r_b$ which specifies the outer 
boundary, consider the following assumption on the 
initial data (one of compact support):
\begin{equation} \label{eq:IDassume}
\Psi_\ell(0,r) = 0 = (\partial_t\Psi_\ell)(0,r),\qquad
r > r_b - \delta,\qquad \text{for any small }\delta > 0.
\end{equation}
Provided that \eqref{eq:IDassume} holds, 
in the region $r \geq r_b$ Laplace transformation of 
an {\em outgoing} mode $\Psi_{\ell}(t,r)$ yields 
\begin{equation}\label{eq:ellmult_fd}
\widehat{\Psi}_\ell(s,r) = a(s)s^\ell e^{-sr} W_\ell(sr),\quad
        W_\ell(z) = \sum_{k=0}^\ell \frac{c_{\ell k}}{z^k},
\end{equation}
where $z=sr$ and 
\begin{equation}\label{eq:profile_fd}
      a(s) \equiv e^{sr}\int_0^\infty e^{-st}f(t-r)dt
             =  \int_{-r_b}^\infty e^{-su}f(u) du.
\end{equation}
Notice that $a(s)$ is indeed independent of $r$. 
For the quadrupole case, we have
\begin{equation}\label{eq:ell2mult_fd}
\widehat{\Psi}_2(s,r) = a(s)s^2 e^{-sr} W_2(sr),\qquad
        W_2(z) = 1 + \frac{3}{z}+\frac{3}{z^2}
\end{equation}
[compare this expression with Eq.~\eqref{eq:ell2mult_td}].

To obtain \eqref{eq:ellmult_fd} and \eqref{eq:profile_fd}, 
we have used \eqref{eq:IDassume} as follows. First 
consider the Laplace transform of $f^{(\ell-k)}(t-r)$ which 
appears in \eqref{eq:defpsil}. Repeated integration by parts 
generates $t=0$ boundary terms of the form $f^{(\ell-k-p)}(-r)$ 
for $1 \leq p \leq \ell - k$. All such terms vanish, as can be
shown by the following identity:
\begin{equation}\label{eq:ffromPsi}
f(t-r) = (-1)^\ell \frac{2^\ell}{(2\ell)!}
\left[r^2\left(\partial_t
+ \partial_r\right)\right]^\ell\Psi_\ell(t,r).
\end{equation}
Since the initial data $\Psi_\ell$ and $\partial_t\Psi_\ell$
vanishes on an open {\em spatial} neighborhood of the spatial
point with coordinate $r \geq r_b$, in fact $\Psi_\ell$ vanishes in an 
open {\em spacetime} neighborhood of the spacetime point 
with coordinates $(0,r)$. Therefore, {\em all} derivatives of 
$\Psi_\ell$ vanish in the same neighborhood, which implies 
$f^{(\ell-k-p)}(-r) = 0$. This implication stems from repeated 
differentiation of \eqref{eq:ffromPsi} with $t=0$ enforced 
afterward.

The previous argument establishes that
\begin{equation}
\int^\infty_0 e^{-st}f^{(\ell-k)}(t-r)dt
= s^{\ell-k} e^{-sr}\int^\infty_{-r} e^{-su}f(u)du.
\end{equation}
The lower limit $-r$ of integration can now be replaced with $-r_b$. 
Indeed, \eqref{eq:IDassume} implies that $\Psi(t,r)=0$ 
for $0 \leq t < \delta + (r-r_b)$ and $r\geq r_b$. Using 
\eqref{eq:ffromPsi}, we conclude that $f(u) = 0$ for $-r \leq
u \leq -r_b$. Therefore, the right-hand side of the last equation 
is $a(s) s^{\ell-k} e^{-sr}$.

\subsubsection{Radiation boundary conditions for flatspace multipoles}
We continue to derive expressions for $r \geq r_b$
where the assumption \eqref{eq:IDassume}
of compact support holds. The explicit expression \eqref{eq:ellmult_fd} 
for $\widehat{\Psi}_\ell(s,r)$ determines an exact
frequency-domain boundary condition
\begin{equation} \label{eq:SommerfeldReshat}
       s\widehat{\Psi}_\ell(s,r) + \partial_r\widehat{\Psi}_\ell(s,r)
      = \frac{1}{r} \widehat{\Omega}_\ell(s,r)\widehat{\Psi}_\ell(s,r),
\end{equation}
where the {\em frequency-domain radiation kernel} 
$\widehat{\Omega}_\ell(s,r)$ 
defines the Sommerfeld residual. Indeed, the operator on the left-hand 
of \eqref{eq:SommerfeldReshat} corresponds to the Sommerfeld operator 
$\partial_t + \partial_r$ in the time-domain. 
If $\ell = 0$, then $\widehat{\Omega}_\ell(s,r) = 0$;
otherwise a simple computation based on 
Eqs.~(\ref{eq:ellmult_fd},\ref{eq:SommerfeldReshat}) shows
that the frequency-domain kernel is given by (with the prime 
indicating differentiation in argument): 
\begin{equation} \label{eq:Omegahat}
\widehat{\Omega}_\ell(s,r)
\equiv sr \frac{W'_\ell(sr)}{W_\ell(sr)}
= \sum_{k=1}^\ell \frac{b_{\ell,k}/r}{s-b_{\ell,k}/r},
\end{equation}
where $\{b_{\ell,k} : 1 \leq k \leq \ell \}$ are the roots 
of $W_\ell(z)$, all of which are simple.\footnote{\label{fn:MacDonald}
The last equality also follows from the identity
\begin{equation}
W_\ell(z) = \sqrt{\frac{2 z}{\pi}}e^z K_{\ell + 1/2}(z),
\end{equation}
showing that the $b_{\ell,k}$ are also the roots of the
half-integer MacDonald function $K_{\ell+1/2}(z)$, which
are simple and lie in the left-half plane \cite{Olver,Watson}. 
The appearance of $K_{\ell+1/2}(z)$ 
may have been anticipated; indeed, the modified Bessel equation 
arises when finding separable solutions to the Laplace 
transformed flatspace radial wave equation \eqref{eq:flatradial}.
\label{fn:MacDonald}}
For the quadrupole case
\begin{equation} \label{eq:Omegahat2}
\widehat{\Omega}_2(s,r) 
= \frac{z_+/r}{s-z_+/r} + \frac{z_-/r}{s-z_-/r},
\quad
z_\pm = -\frac{3}{2}\pm \mathrm{i}\frac{\sqrt{3}}{2},
\end{equation}
where $z_+ = b_{2,1}$ and $z_- = b_{2,2}$ solve $W_2(z_\pm) = 0$. 

The time-domain RBC is the inverse Laplace transform of
\eqref{eq:SommerfeldReshat}, i.e.~the Laplace 
convolution \cite{GroteKeller1,GroteKeller2,Sofronov1,Sofronov2,AGH}
\begin{equation} \label{eq:SommerfeldRes}
      \partial_t\Psi_\ell + \partial_r\Psi_\ell = \frac{1}{r}\int_0^t
\Omega_\ell(t-t',r)\Psi_\ell(t',r)dt',\quad
 \Omega_\ell(t,r)= \sum_{k=1}^\ell \frac{b_{\ell,k}}{r}
\exp\left(\frac{b_{\ell,k}t}{r}\right).
\end{equation}
Subject to our assumption \eqref{eq:IDassume}
of compact support, the outgoing 
multipole [$\epsilon = 1$ in Eq.~\eqref{eq:defpsil}] obeys 
Eq.~\eqref{eq:SommerfeldRes} exactly, as can also be shown via 
direct calculation using repeated integration by parts;
see Appendix \ref{sec:NoLaplace}. If, on the other hand, 
assumption \eqref{eq:IDassume} does not hold, then 
Eq.~\eqref{eq:SommerfeldRes} is violated, but only by terms 
which decay exponentially fast in $t$; again, see Appendix 
\ref{sec:NoLaplace}.

\subsubsection{Asymptotic waveform evaluation and teleportation
for flatspace multipoles} 
For a generic outgoing solution, it is possible to recover 
the profile function $f(t-r)$ and asymptotic waveform 
\begin{equation}
\Psi_\ell(t,r) \sim f^{(\ell)}(t-r),\qquad r\rightarrow\infty,
\end{equation}
via data recorded solely at a finite and {\em fixed} 
radial location, again taken as $r = r_b$.
Let us consider the $\ell = 2$ case as a concrete example.
Generalization to higher $\ell$ is straightforward. Equation
\eqref{eq:ell2mult_td} suggests that we solve the ODE initial 
value problem
\begin{equation}\label{eq:extractionODE}
y'' + \frac{3}{r} y' + \frac{3}{r^2}y 
= \Psi_2(t,r), \qquad
y(0) = 0 = y'(0),\qquad
(\ell = 2 \text{ problem})
\end{equation}
in which case $f(u) = y(u+r)$. 

In a pioneering series of 
papers \cite{AE1,AE2}, Abrahams and Evans showed how 
the above procedure carries over to the theory of gravitational 
multipoles for general relativity linearized about flat 
spacetime. We now re-examine the basic 
idea behind Abrahams-Evans AWE from the standpoint of 
Laplace convolution, and will consider two kernels:
one $\Theta_\ell$ for evaluation of the underlying 
function $f(u)$ and another $\Phi_\ell$ more suited for 
evaluation of the waveform $f^{(\ell)}(u)$. Our implementations
have mostly relied on the $\Phi_\ell$ kernel.

Continuing with the $\ell = 2$ example, we introduce a
{\em frequency-domain profile evaluation kernel} 
$\widehat{\Theta}_2(s,r)$ tailored to satisfy
\begin{equation}
\widehat{\Theta}_2(s,r)\widehat{\Psi}_2(s,r) 
= a(s) e^{-sr} = \widehat{y}(s).
\end{equation}
That is, the product of $\widehat{\Theta}_2(s,r)$ and 
$\widehat{\Psi}_2(s,r)$ is $\int_0^\infty e^{-st}f(t-r)dt$ 
[cf.~Eq.~\eqref{eq:profile_fd}]. Comparison with 
\eqref{eq:ell2mult_fd} immediately shows that
\begin{equation}\label{eq:flatspaceTheta}
      \widehat{\Theta}_2(s,r) = \frac{1}{s^2 W_2(sr)} =
      \frac{\mathrm{i}r}{\sqrt{3}}
      \left[
      \frac{1}{s-z_-/r} - \frac{1}{s-z_+/r}
      \right].
\end{equation}
The corresponding time-domain profile evaluation kernel is 
\begin{equation}\label{eq:FLAT_TDK_2}
      \Theta_2(t,r) = \frac{\mathrm{i}r}{\sqrt{3}}
      \left[
      \exp\left(\frac{z_-t}{r}\right)-
      \exp\left(\frac{z_+t}{r}\right)
      \right],
\end{equation}
and $y(t) = (\Theta_2(\cdot,r) * \Psi_2(\cdot,r))(t)$ solves the 
Abrahams-Evans initial value problem \eqref{eq:extractionODE}. 
Essentially the same arguments show that the order-$\ell$
profile evaluation kernel is
\begin{equation}
   \widehat{\Theta}_\ell(s,r) = 
\frac{1}{s^\ell W_\ell(sr)} .
\end{equation}
Despite appearances, the kernel is regular at $s = 0$. For example, 
$1/W_2(sr) \sim (sr)^2/3$, and in general
$1/W_\ell(sr) \sim (sr)^\ell/c_{\ell\ell}$, as $s\rightarrow 0$. 

Direct evaluation of the asymptotic waveform is also possible. 
{\em Teleportation} by a positive shift $r_2-r_1$ means conversion of 
$\Psi(t,r_1)$ to $\Psi(t+(r_2-r_1),r_2)$, and it might 
correspond to a small finite shift $r_2-r_1$. However, when $r_2$ 
is suitably large, we write $r_\infty$ for $r_2$ and view 
teleportation as an AWE procedure (in which case typically
$r_1 = r_b$, and it is the boundary waveform $\Psi(t,r_b)$ which 
is teleported). Teleportation is accomplished with a 
{\em frequency-domain teleportation kernel} 
\begin{equation}\label{eq:freqPhidef}
\widehat{\Phi}_\ell(s,r_1,r_2)
         = -1 + \frac{W_\ell(sr_2)}{W_\ell(sr_1)}
\end{equation}
rigged to satisfy
\begin{equation}
e^{s(r_2-r_1)}\widehat{\Psi}_\ell(s,r_2) =
\widehat{\Phi}_\ell(s,r_1,r_2)\widehat{\Psi}_\ell(s,r_1)
+ \widehat{\Psi}_\ell(s,r_1).
\end{equation}
We have included the $-1$ factor in \eqref{eq:freqPhidef} to 
ensure that $\widehat{\Phi}_\ell(s,r_1,r_2)$ has a 
well-defined inverse Laplace transform $\Phi_\ell(t,r_1,r_2)$.
In the time domain we recover the desired property
\begin{equation}\label{eq:teleport}
\Psi_\ell(t+(r_2-r_1),r_2) =
(\Phi_\ell(\cdot,r_1,r_2)*\Psi_\ell(\cdot,r_1))(t) + 
\Psi_\ell(t,r_1).
\end{equation}
Adjusting for the $(r_2 - r_1)$ time delay, this
formula allows for conversion of the signal at $r_1$
to the signal at $r_2$. Since $r_2 \leq \infty$,
this method can also be used for evaluation
of the asymptotic waveform $f^{(\ell)}(u)$. We refer
to the $r_2 = \infty$ case $\widehat{\Phi}_\ell(s,r_1,\infty)$ 
as the {\em frequency-domain waveform evaluation kernel}.

The relationship between RBC and 
AWE/teleportation kernels
is a key insight of this paper, and the one which is exploited 
to numerically construct AWE/teleportation kernels.
For example, the profile evaluation kernel can be written as
\begin{equation}\label{eq:flatspaceThetaOmega}
   \widehat{\Theta}_\ell(s,r) = 
\frac{1}{s^\ell}
 \underbrace{\exp\left[\int_r^\infty
\frac{\widehat{\Omega}_\ell(s,\eta)}{\eta} d\eta \right]}_{1/W_\ell(sr)}.
\end{equation}
That the underbraced quantity is indeed $1/W_\ell(sr)$ follows
easily from the identity $\eta^{-1}\widehat{\Omega}_\ell(s,\eta) 
= \partial_\eta \log W_\ell(s\eta)$, 
that is essentially the definition \eqref{eq:Omegahat}. 
The integration in \eqref{eq:flatspaceThetaOmega} can of course be 
carried out, recovering \eqref{eq:flatspaceTheta} for the
$\ell = 2$ case; however, when considering similar expressions 
for blackhole perturbations at least some of the integration 
will be performed by numerical quadrature.
Similarly, one can express the teleportation kernel through
\begin{equation}\label{eq:flatspacePhi}
\widehat{\Phi}_\ell(s,r_1,r_2)
         = -1 + \underbrace{\exp\left[\int_{r_1}^{r_2}
           \frac{\widehat{\Omega}_\ell(s,\eta)}{\eta}d\eta\right]}_{
           W_\ell(sr_2)/W_\ell(sr_1)}.
\end{equation}

In Section~\ref{subsec:BH_perts} we introduce the analogous
kernels for waveform teleportation in the
Regge-Wheeler and Zerilli formalisms. As mentioned,
we have mostly used the $\Phi_\ell$ kernels.

\subsubsection{Efficiency and storage} \label{sec:Eff}
Here we comment on RBC and AWE for the ordinary 3+1 wave equation
from the standpoint of efficiency and storage 
as $\ell \rightarrow \infty$, both summarizing
known results for RBC \cite{AGH} and considering these issues 
for AWE. Let $\lambda$ represent a
characteristic wavelength, say determined by the initial
data or inputted boundary conditions.
For numerical evolution to a fixed final time $T$, 
an implementation of the {\em exact} flatspace RBC 
\eqref{eq:SommerfeldRes} (with a kernel comprised 
of $\ell$ exponentials) has the following 
work and storage requirements:
\begin{equation}\label{eq:exactRBCscalings}
\mathrm{Work}_\mathrm{exactRBC} = O(\lambda^{-1}\ell T), \qquad
\mathrm{Storage}_\mathrm{exactRBC} = O(\ell).
\end{equation}
These scalings are deduced from the cost of integrating $\ell$ 
ODE of the form \eqref{eq:boundaryODE} with
approximately $\lambda^{-1} T$ timesteps.

A {\em spatially and temporally resolved} numerical 
integration (with arbitrary boundary conditions) of 
Eq.~\eqref{eq:flatradial} on a radial domain of fixed 
size corresponds to the following work and storage scalings: 
$O(\lambda^{-2}T)$ and $O(\lambda^{-1})$. Indeed,
a resolved spatial discretization of Eq.~\eqref{eq:flatradial}
yields a coupled system of approximately $\lambda^{-1}$ ODE.
As more spatial/temporal resolution is typically required for 
large $\ell$ solutions, it is reasonable to view 
$\lambda^{-1} \simeq \ell$, in which case the scalings 
for the interior solver are comparable to 
\eqref{eq:exactRBCscalings}.
However, implementation of the exact RBC is still preferable to
choosing the computational domain so large that the outer 
boundary is casually disconnected from the wordline of an 
interior ``detector". Spatial discretization after 
such domain enlargement yields $\lambda^{-1}T$ 
coupled ODE, whence $O(\lambda^{-2}T^2)$ and $O(\lambda^{-1}T)$ for
the work and storage.

{\em Kernel compression} yields a more efficient implementation
of RBC. As proven in Ref.~\cite{AGH}, the 
kernel $\widehat{\Omega}_\ell(s,r)$ admits a rational 
approximation\footnote{The 
$\gamma_{\ell,n}$ and $\beta_{\ell,n}$ appearing in
the approximate (frequency-domain) flatspace kernel
\eqref{eq:compressedFLTkern} are different
than the similar parameters appearing in the 
approximate (time-domain) blackhole kernel
\eqref{eq:compressedBHkern}. Here 
$\gamma_{\ell,n}$ and $\beta_{\ell,n}$ {\em do not} 
depend on $r$, whereas the parameters in 
\eqref{eq:compressedBHkern} {\em do} depend on the 
(rescaled) radius. We use similar notations for
the flatspace and blackhole cases, hoping this
practice does not cause confusion.}
\begin{equation}\label{eq:compressedFLTkern}
\widehat{\Xi}_\ell(s,r) = 
\sum_{n=1}^d \frac{\gamma_{\ell,n}/r}{s-\beta_{\ell,n}/r},
\qquad
\sum_{s\in \mathrm{i}\mathbb{R}}
\left|\frac{\widehat{\Omega}_\ell(s,r)-\widehat{\Xi}_\ell(s,r)}{
\widehat{\Omega}_\ell(s,r)}\right| < \varepsilon,
\end{equation}
where $\varepsilon$ is a prescribed tolerance and the number 
of approximating poles scales like \cite{AGH}
\begin{align}\label{eq:AGHdscaling}
d = O\big(\log\nu\log(1/\varepsilon)
+\log^2\nu+\nu^{-1}\log^2(1/\varepsilon)\big)
\end{align}
as $\nu = \ell+1/2 \rightarrow \infty$ and $\varepsilon
\rightarrow 0^{+}$. The frequency domain bound in 
\eqref{eq:compressedFLTkern} implies a long-time 
bound on the relative convolution error
in the time-domain, see Appendix \ref{sec:ErrorEst}. Since $d$ 
grows sublinearly in $\ell$ and $1/\varepsilon$, the approximation 
$\widehat{\Xi}_\ell(s,r)$ [likewise its inverse Laplace transform 
$\Xi_\ell(t,r)$] is called a {\em compressed kernel}. An 
implementation of Laplace convolution RBC based on compressed 
kernels $\Xi_\ell(t,r)$ scales like
\begin{equation}\label{eq:RBCscalings}
\mathrm{Work}_\mathrm{compressedRBC} = O(\lambda^{-1}dT),
\qquad \mathrm{Storage}_\mathrm{compressedRBC} = O(d),
\end{equation}
with clear performance in the large-$\ell$ limit.

The proof of \eqref{eq:AGHdscaling} relies on the large-$\ell$
asymptotics \cite{AGH,Olver} of the roots 
$\{b_{\ell,k}: k = 1,\dots,\ell\}$ of $K_{\ell + 1/2}(z)$. 
Precisely, as $\ell \rightarrow \infty$ the scaled roots 
$b_{\ell,k}/(\ell+1/2)$ accumulate on a curve $\mathcal{C}$
given by \cite{AGH,Olver}
\begin{equation}\label{eq:curveC}
z(\lambda) = -\sqrt{\lambda^2 -\lambda \tanh\lambda}
\pm \mathrm{i} \sqrt{\lambda \coth\lambda-\lambda^2},
\qquad
\lambda \in [0,\lambda_0],\qquad
\tanh\lambda_0 = 1/\lambda_0.
\end{equation}
Since the pole locations appearing in {\em both} the exact 
flatspace RBC and AWE kernels are
$\{b_{\ell,k}/r: k = 1,\dots,\ell\}$,
we conjecture that an implementation 
of AWE based on kernel compression 
{\em formally} satisfies the scalings \eqref{eq:RBCscalings}. 
However, we are unsure if these scalings hold in practice. 

\begin{figure}
\includegraphics[clip=true,width=16cm,trim=0 4.25cm 0 0cm]
{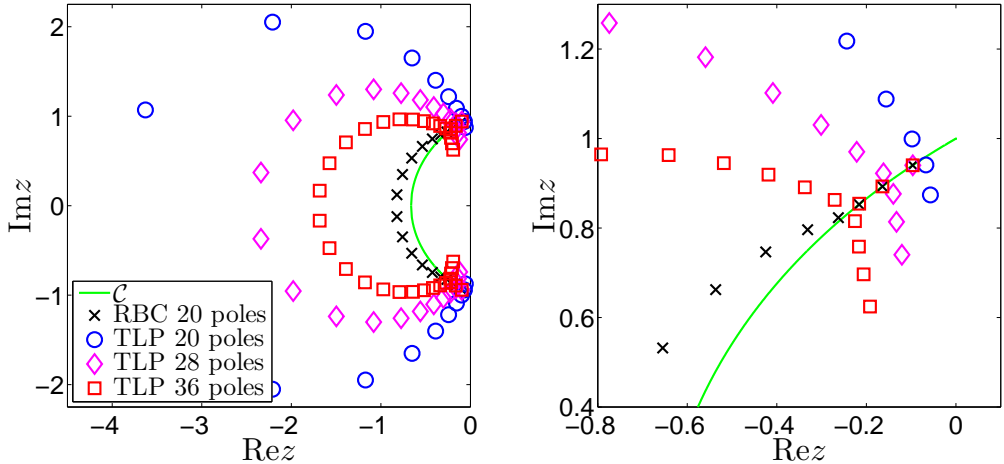}
\caption{{\sc Scaled pole locations for compressed kernels.}
Here TLP means {\em teleportation}, and the curve $\mathcal{C}$
is described by the parameterization $z(\lambda)$ given in 
\eqref{eq:curveC}. In the blow-up plot there is a 
cross-diamond-square coalescence on $\mathcal{C}$ (near its
right end). See the text for further explanation.}
\label{fig:PolesFlatspaceRBCandTLPell64}
\end{figure}

As a nascent investigation, we consider compressed kernels 
for $\ell = 64$ flatspace RBC and teleportation. 
Figure \ref{fig:PolesFlatspaceRBCandTLPell64} plots 
{\em scaled} pole locations for a 20-pole compressed 
kernel $\widehat{\Xi}_{64}(s,15)$ which approximates
$\widehat{\Omega}_{64}(s,15)$ and for 20, 28, and 
36-pole versions of a compressed kernel 
$\widehat{\Xi}{}^E_{64}(s,15,240)$ which approximates 
$\widehat{\Phi}_{64}(s,15,240)$. Here we have scaled 
all pole locations by a factor $r/(\ell+1/2) = 15/64.5$ in 
order to plot them relative to the curve $\mathcal{C}$, 
on which the actual scaled zeros $b_{64,k}/64.5$ lie
(at least to the eye). Figure \ref{fig:PolesFlatspaceRBCandTLPell64}
shows that, compared with poles for compressed teleportation 
kernels, the poles for the compressed RBC kernel lie much closer
to $\mathcal{C}$. Nevertheless, for both compressed RBC and 
teleportation kernels as the number of approximating poles 
increases (corresponding to a smaller tolerance $\varepsilon$),
more of the approximating poles ``lock on" to $\mathcal{C}$.
This behavior is evident in the right blow-up plot, where for 
20, 28, and 32-pole compressed teleportation kernels, we respectively
find 0(circles), 1(diamond), and 3(squares) ``locked-on" poles. 
Moreover, at least to the eye, these correspond to ``locked-on" 
poles (crosses) for the compressed RBC kernel. Sec.~\ref{sec:EffBH} 
briefly discusses these issues for the gravitational case.

\subsection{Blackhole perturbations} \label{subsec:BH_perts}

We consider the following
rescaled versions of the generic master 
equation \eqref{eq:RWZeqn} (retaining 
the same stem letter $V$ for the potentials):
\begin{equation}\label{eq:RWZeqn_rho}
       \frac{\partial^2\Psi_\ell}{\partial \tau^2}
                  -\frac{\partial^2\Psi_\ell}{\partial \rho_*^2} +
                   V^\mathrm{RW,Z}_\ell(\rho)\Psi_\ell = 0,
\end{equation}
here in terms of rescaled coordinates
\begin{equation}
\rho = r/(2M),\quad
                    \tau = t/(2M),\quad
        \rho_* = \rho + \log(\rho -1).
\end{equation}
Expressions for the Regge-Wheeler 
$V^\mathrm{RW}_\ell(\rho)$ and Zerilli 
$V^\mathrm{Z}_\ell(\rho)$ 
potentials are given in Eqs.~(3) and (4) of 
Ref.~\cite{Lau3} (expressions in 
terms of $r$ rather than $\rho$ are given in 
\cite{FHL}).
The formulas we present here hold
for both formalisms, and have been drawn from
Refs.~\cite{Lau1,Lau2,Lau3}. As before in our analysis
of the flatspace radial wave equation, here we also
suppress the azimuthal index $m$ on $\Psi_{\ell}$.

\subsubsection{Structure of outgoing solutions}
With $\sigma = 2Ms$ the rescaled Laplace frequency, 
formal Laplace transformation of \eqref{eq:RWZeqn} yields
\begin{equation}\label{eq:RWZeqn_Laplace}
- \frac{d^2\widehat{\Psi}_\ell}{d \rho_*^2} 
+ V^\mathrm{RW,Z}_\ell(\rho)\widehat{\Psi}_\ell + 
\sigma^2\widehat{\Psi}_\ell = 0,
\end{equation}
Outgoing solutions of the last equation can be expressed 
as an asymptotic series\footnote{\label{fn:asymptotic_series}
The coefficients $d_{\ell,k}(\sigma)$ defining
the asymptotic series $W_\ell(z,\sigma) \sim \sum_{k=0}^\infty
d_{\ell,k}(\sigma)z^{-k}$ are respectively defined by three-term
and five-term recursion relations in the Regge-Wheeler
and Zerilli formalisms~\cite{Lau1,Lau2,Lau3}.} about $\rho=\infty$,
\begin{equation} \label{eq:OutgoingBH}
\widehat{\Psi}_\ell(\sigma,\rho) = a(\sigma)\sigma^\ell
e^{-\sigma\rho_*} W_\ell(\sigma\rho,\sigma),\qquad
W_\ell(\sigma\rho,\sigma) 
\mathop{\sim}_{\scriptstyle \rho\rightarrow\infty} 1
\end{equation}
[cf.~Eq.~\eqref{eq:ellmult_fd} for a flatspace
multipole].
Notice that $W_\ell(\sigma\rho,\sigma) = W_\ell(sr,2Ms)$,
and so {\em formally} the flatspace expression $W_\ell(sr) = 
W_\ell(sr,0)$.

\subsubsection{Radiation boundary conditions}
We again assume initial data of compact support, namely that
$\Psi_\ell(0,\rho) = 0 = (\partial_\tau\Psi_\ell)(0,\rho)$ for
$\rho > \rho_b - \delta$, where $\rho_b$ specifies the outer
boundary. We now work with any $\rho \geq \rho_b$, in terms 
of which exact radiation conditions satisfied by a generic 
asymptotically outgoing multipole \eqref{eq:OutgoingBH}
have the following frequency-domain and time-domain 
forms~\cite{Lau1,Lau2,Lau3}:
\begin{equation}\label{eq:BH_RBC}
            \sigma \widehat{\Psi}_\ell
+ \partial_{\rho_*}\widehat{\Psi}_\ell
= \frac{1}{\rho}\left(1-\frac{1}{\rho}\right)
   \widehat{\omega}_\ell\widehat{\Psi}_\ell
\iff
     \partial_\tau \Psi_\ell
+ \partial_{\rho_*}\Psi_\ell
= \frac{1}{\rho}\left(1-\frac{1}{\rho}\right)
\omega_\ell * \Psi_\ell,
\end{equation}
where the frequency domain radiation kernel is 
\begin{equation}
\widehat{\omega}_\ell(\sigma,\rho) \equiv 
\sigma\rho
\frac{W'_\ell(\sigma\rho,\sigma)}{W_\ell(\sigma\rho,\sigma)},
\end{equation} 
with the prime denoting differentiation in the first argument.
\begin{figure}
\includegraphics[clip=true,width=14.0cm]{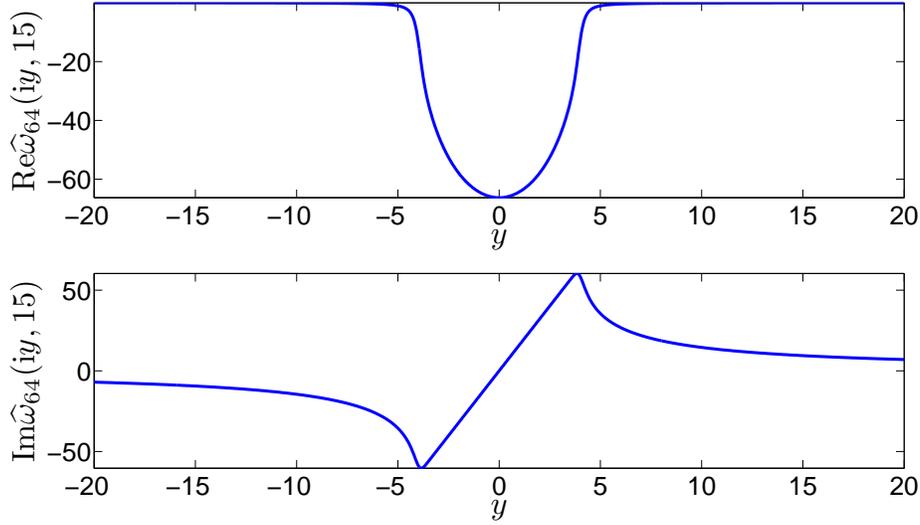}
\caption{{\sc Profiles for an $\rho = 15$, $\ell = 64$ Regge-Wheeler
frequency domain RBC kernel.}
}
\label{fig:RWell64}
\end{figure}

Refs.~\cite{Lau1,Lau3} have argued that the kernel
$\widehat{\omega}_\ell(\sigma,\rho)$ has the following 
``sum of poles" representation:\footnote{This terminology 
is suggestive only, since in complex analysis poles are 
{\em isolated singularities}. Therefore, the integral term 
$\widehat{\omega}_\ell^\mathrm{cut}(\sigma,\rho)$ appearing 
in \eqref{eq:omegaSOP} is not, strictly speaking, a 
``continuous distribution of poles."}
\begin{equation}\label{eq:omegaSOP}
\widehat{\omega}_\ell(\sigma,\rho)
= \widehat{\omega}_\ell^\mathrm{pole}(\sigma,\rho) + 
\widehat{\omega}_\ell^\mathrm{cut}(\sigma,\rho)
\equiv \sum_{k=1}^{N_\ell}
\frac{\sigma_{\ell,k}'(\rho)}{\sigma - \sigma_{\ell,k}(\rho)}
-\frac{1}{\pi}\int_0^\infty 
\frac{f_\ell(\chi;\rho)}{\sigma+\chi}d\chi,
\end{equation}
where $f_\ell(\chi;\rho) \equiv 
\mathrm{Im}\widehat{\omega}_\ell(\chi e^{\mathrm{i}\pi},\rho)$ 
and the $\sigma_{\ell,k}(\rho)$ are simple roots of 
$W_\ell(\sigma\rho,\sigma)$ (analogous to the roots
$b_{\ell,k}/r$ of $W_\ell(sr)$ in the flatspace case).
At least for $\rho \geq 15$, the integer $N_\ell =
\ell$ or $\ell + 1$, when $\ell$ is respectively even 
or odd \cite{Lau1,Lau3}. The origin of the extra root,
relative to the flatspace case, in the odd-$\ell$ case
is discussed in Ref.~\cite{Lau3}.
\begin{figure}
\includegraphics[clip=true,width=14.0cm]{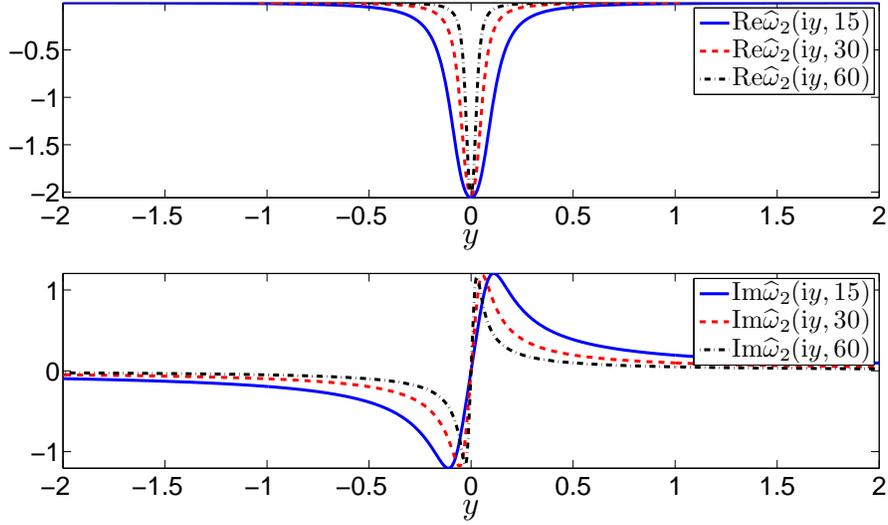}
\caption{{\sc Profiles for $\ell = 2$ Regge-Wheeler 
frequency domain RBC kernels.}
As $\rho$ increases these profiles shrink towards the origin 
(the same phenomena occurs for the flatspace RBC kernels).
}
\label{fig:RWell2_shrink}
\end{figure}

Insofar as numerical implementation is concerned, a key 
requirement is the ability to evaluate the profiles
$\mathrm{Re}\widehat{\omega}_\ell(\mathrm{i}y,\rho)$ and
$\mathrm{Im}\widehat{\omega}_\ell(\mathrm{i}y,\rho)$ for
$y\in\mathbb{R}$. These evaluations are along
the imaginary $\sigma$-axis, typically the inversion
contour for the inverse Laplace transform. Accurate 
methods for such evaluation have been described in \cite{Lau1}.
In fact, these methods are {\em not} based on the sum-of-poles
representation \eqref{eq:omegaSOP}, but this issue
is of no concern here. For example, the profiles 
for an $\rho = 15$, $\ell = 64$ Regge-Wheeler kernel are 
shown in Fig.~\ref{fig:RWell64}. As $\rho$ increases 
these profiles ``shrink" towards the origin (as do the 
corresponding flatspace profiles), and this phenomena is
documented in Fig.~\ref{fig:RWell2_shrink}.
With the ability to numerically 
generate the profiles 
$\mathrm{Re}\widehat{\omega}_\ell(\mathrm{i}y,\rho)$
and $\mathrm{Im}\widehat{\omega}_\ell(\mathrm{i}y,\rho)$, 
we are then able to construct approximate kernels 
via Alpert-Greengard-Hagstrom (AGH) compression \cite{AGH}. 
Here a compressed kernel is a sum of simple poles,
\begin{equation}\label{eq:compressedK}
\widehat{\xi}_\ell(\sigma,\rho) \equiv
\sum_{q=1}^d 
\frac{\gamma_{\ell,q}(\rho)}{\sigma-\beta_{\ell,q}(\rho)}
\simeq
\widehat{\omega}_\ell(\sigma,\rho),
\qquad 
\mathrm{Re}\beta_{\ell,q}(\rho) < 0, \forall q,
\end{equation}
where the approximation $\widehat{\xi}_\ell(\sigma,\rho)$ 
satisfies
\begin{equation}\label{eq:superr}
\left|\widehat{\omega}_\ell(\sigma,\rho)
- \widehat{\xi}_\ell(\sigma,\rho)
\right| <
\varepsilon \left|\widehat{\omega}_\ell(\sigma,\rho)\right|,
\qquad
\sigma \in
\mathrm{i}\mathbb{R},
\end{equation}
with $\varepsilon$ a prescribed tolerance. The number
$d$ clearly depends on $\varepsilon$ and $\ell$, 
and the numbers $\beta_{\ell,q}$ and $\gamma_{\ell,q}$ 
depend both on the boundary radius $\rho$ (as indicated) and 
on $\ell$ (the dependence on which we have restored here). 
The modifier {\em compressed} in the description 
of $\widehat{\xi}_\ell(\sigma,\rho)$ is apt. 
Indeed, as described in Sec.~\ref{sec:Eff}, for the ordinary 
wave equation 
the exact frequency domain kernel admits a similar rational
approximation with $d$ scaling as \eqref{eq:AGHdscaling}.
Similar scaling has been observed empirically for approximations
$\widehat{\xi}_\ell(\sigma,\rho)$ of blackhole
kernels $\widehat{\omega}_\ell(\sigma,\rho)$ \cite{Lau1}.

\begin{figure}
\includegraphics[trim=0cm 4cm 0cm 4cm, clip=true,width=12.0cm]{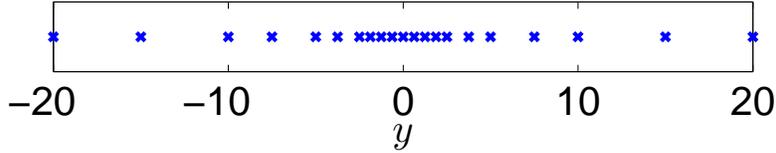}
\caption{{\sc $y$-grid to evaluate the profiles 
$\mathrm{Re}\widehat{\omega}_2(\mathrm{i}y_j,\rho)$
and $\mathrm{Im}\widehat{\omega}_2(\mathrm{i}y_j,\rho),\forall j$.}
This grid has 21 points, 4 adaptive levels, and 9 points at the bottom level.}
\label{fig:ygrid}
\end{figure}

Algorithm \ref{tab:AGHcomp} summarizes
our implementation of AGH compression (see Ref.~\cite{Lau2}
for a complete description). Let us further comment on 
Algorithm \ref{tab:AGHcomp}, with numbers appropriate for 
$\ell = 2$. Typically $y_\mathrm{max} = 300/\rho$ for step 1. 
For step 2 we have typically chosen 10 to 20 adaptive 
levels centered around the origin with 65 points at the 
bottom level, about $10^3$ grid points $y_j$ in all.
Fig.~\ref{fig:ygrid} depicts an example $y$-grid.
For step 3 the evaluation at each $y_j$ requires ODE integration
in the complex plane with upwards of $10^{10}$ floating
point operations in double precision (more in quad precision).
Step 5 is a confirmation step meant to verify \eqref{eq:superr}.
Ideally, this confirmation takes place with a 
much larger $y$-window than  $[-y_\mathrm{max},y_\mathrm{max}]$, and
on a different (dense and uniform) $y$-grid. This step involves 
further ODE integration and is therefore as or more expensive than
step 3.
\begin{algorithm}[H]
\caption{{\sc Steps for compressing an RBC kernel.}\label{alg:compressRBC} }
\label{tab:AGHcomp}
\vspace{5pt}
INPUT: $\ell$, $\rho = \rho_b$, $d$ 
       (orbital index, dimensionless boundary radius,
        desired number of poles) \\
OUTPUT:  $\{ \beta_{\ell,q}(\rho), 
            \gamma_{\ell,q}(\rho): 
         q = 1,\dots,d\}$
        (compressed kernel)
\vspace{5pt}
\begin{algorithmic}[1]
\State Choose an approximation window 
       $[-y_\mathrm{max},y_\mathrm{max}]$
       on the $\sigma = \mathrm{i}y$ imaginary axis.
\State Partition $[-y_\mathrm{max},y_\mathrm{max}]$ to 
       form a $y$-grid, typically with mesh refinement 
       at the origin. 
\State
      Numerically evaluate the profiles
      $\mathrm{Re}\widehat{\omega}_2(\mathrm{i}y_j,\rho)$
      and $\mathrm{Im}\widehat{\omega}_2(\mathrm{i}y_j,\rho)$
      on the $y$-grid.
\State
      Compute the numbers $\{\beta_{\ell,q}(\rho),
      \gamma_{\ell,q}(\rho): q = 1,\dots,d\}$
      by AGH compression.
\State 
      Using $\{\beta_{\ell,q}(\rho),
      \gamma_{\ell,q}(\rho): q = 1,\dots,d\}$,
      verify \eqref{eq:superr}. If not verified, 
      repeat with $d \leftarrow d+1$.
\end{algorithmic} 
\end{algorithm}

From the standpoint of implementation, the representation 
\eqref{eq:compressedK} is crucial, since it implies 
that the time-domain convolution can be approximately
evaluated via integration of ODE at the boundary. 
For a typical explicit ODE scheme 
and a sufficiently small time-step, integration of these 
ODE is numerically stable since the relevant poles in 
\eqref{eq:compressedK} lie in the left-half plane. 
Let $\xi_\ell(\tau,\rho)$ be the inverse Laplace transform
of $\widehat{\xi}_\ell(\sigma,\rho)$ with respect to $\sigma$.
Then the approximate time-domain kernel $\Xi_\ell(t,r)
\equiv (1/2M)\xi_\ell(t/(2M),r/(2M))$ appearing in 
\eqref{eq:LaplaceConvApprox} is the inverse Laplace 
transform (with respect to $s$) of
\begin{equation}
\widehat{\Xi}_\ell(s,r) = \sum_{q=1}^d 
\frac{\gamma_{\ell,q}(\rho)}{2M}
\widehat{\Xi}_{\ell,q}(s,r),
\qquad
\widehat{\Xi}_{\ell,q}(s,r) = 
\frac{1}{s-(2M)^{-1}\beta_{\ell,q}(\rho)},
\end{equation}
where, unlike in Section \ref{sec:HowTo},
here $\ell$-dependence has not been suppressed.

\subsubsection{Asymptotic waveform evaluation 
and teleportation}\label{subsec:BHWE}
Similar to before, we introduce two kernels: (i)
one $\theta_\ell$ for evaluation of an underlying profile,
and (ii) another $\phi_\ell$ for AWE/teleportation of the
waveform. The first type of kernel is
defined by [cf.~Eq.~\eqref{eq:flatspaceThetaOmega}]
\begin{equation}
\widehat{\theta}_\ell(\sigma,\rho)
= \frac{1}{\sigma^\ell}\underbrace{\exp\left[\int_\rho^\infty 
\frac{\widehat{\omega}_\ell(\sigma,\eta)}{\eta}
d\eta\right]}_{1/W_\ell(\sigma\rho,\sigma)} \, ,
\end{equation}
and satisfies $\widehat{\theta}_\ell(\sigma,\rho) 
\widehat{\Psi}_{\ell}(\sigma,\rho) = a(\sigma)\exp(-\sigma\rho_*)$, 
as can be seen directly from Eq.~\eqref{eq:OutgoingBH}.
We can not analytically perform the integration here. 
However, the pole part of the kernel can
be exactly integrated to remove the singularity. 
Indeed, we find
\begin{align}
\widehat{\theta}_\ell(\sigma,\rho) =  
\frac{
\exp\left[\int_\rho^\infty
\frac{\widehat{\omega}_\ell^\mathrm{cut}(\sigma,\eta)}{\eta}
d\eta\right]}{
\sigma^\ell\exp\left[-\int_\rho^\infty
\frac{\widehat{\omega}_\ell^\mathrm{pole}(\sigma,\eta)}{\eta}
d\eta\right]}
= \frac{\exp\left[\int_\rho^\infty 
\widehat{\omega}_\ell^\mathrm{cut}(\sigma,\eta)\eta^{-1}d\eta
\right]}{\prod_{k=1}^{N_\ell} 
\big[\sigma-\sigma_{\ell,k}(\rho)\big]}.
\end{align}
Teleportation $(\rho_1 \rightarrow \rho_2 \leq \infty)$ is defined
through the kernel [cf.~Eq.~\eqref{eq:flatspacePhi}]
\begin{align}\label{eq:BHteleportK}
\widehat{\phi}_\ell(\sigma,\rho_1,\rho_2) &  = -1 +
\underbrace{\exp\left[
\int_{\rho_1}^{\rho_2}\frac{\widehat{\omega}_\ell(\sigma,\eta)}{\eta}
d\eta\right]}_{W_\ell(\sigma\rho_2,\sigma)/W_\ell(\sigma\rho_1,\sigma)}.
\end{align}
Adjusting for the $(\rho^*_2 - \rho^*_1)$ time delay, this
kernel allows for conversion of the signal at $\rho_1$
to the signal at $\rho_2$, as it satisfies
\begin{equation}
e^{\sigma(\rho^*_2-\rho^*_1)}\widehat{\Psi}_\ell(\sigma,\rho_2) =
\widehat{\phi}_\ell(\sigma,\rho_1,\rho_2)\widehat{\Psi}_\ell(\sigma,\rho_1)
+ \widehat{\Psi}_\ell(\sigma,\rho_1).
\end{equation}
In the time domain we therefore recover the desired property
\begin{equation}\label{eq:teleport}
\Psi_\ell(\tau+(\rho^*_2-\rho^*_1),\rho_2) =
(\phi_\ell(\cdot,\rho_1,\rho_2)*\Psi_\ell(\cdot,\rho_1))(\tau) + 
\Psi_\ell(\tau,\rho_1).
\end{equation}
Exact evaluation of the asymptotic waveform corresponds to $\rho_2 = \infty$.
\begin{figure}
\includegraphics[clip=true,width=14.0cm]{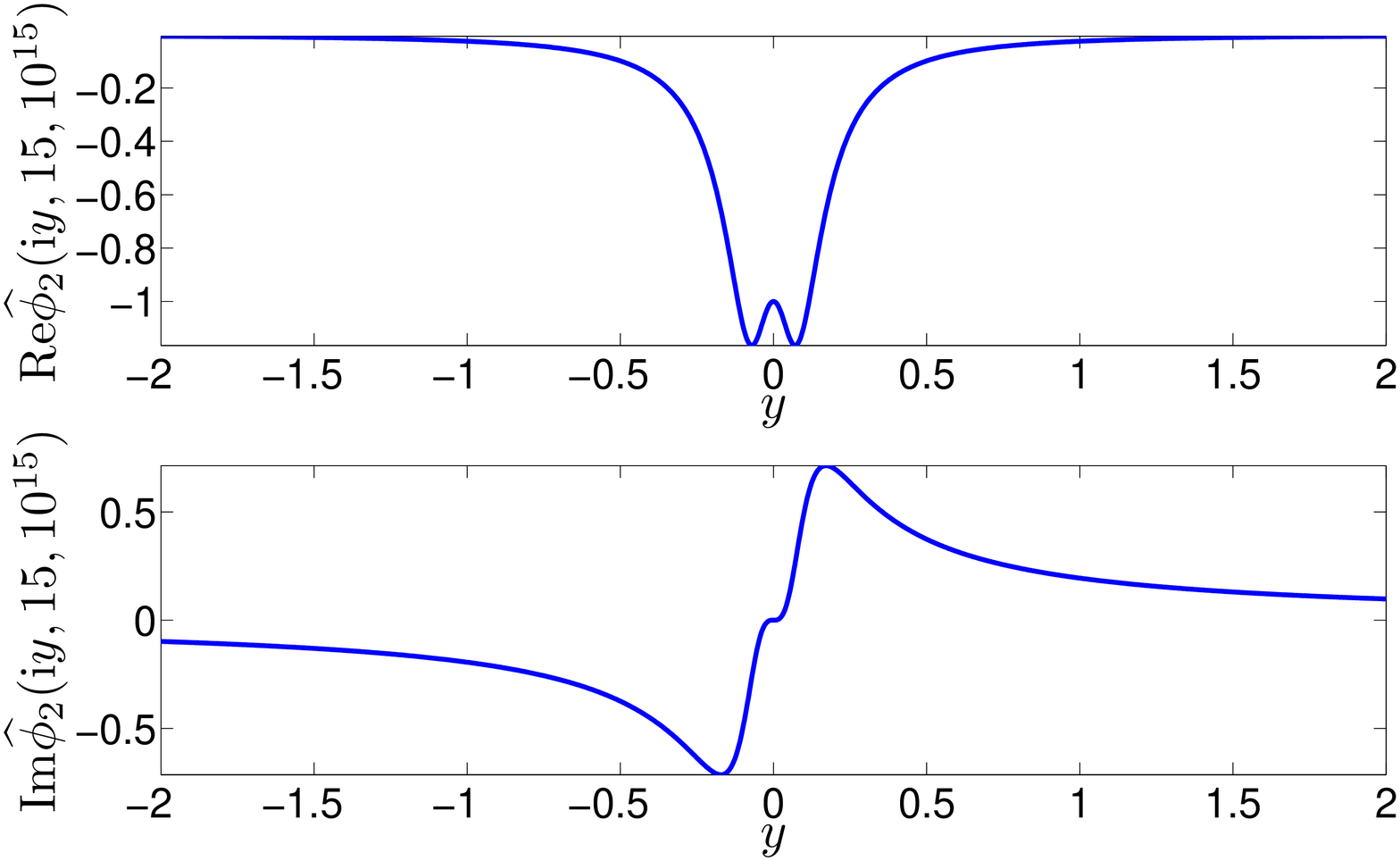}
\caption{Teleportation kernel $\widehat{\phi}_2(\mathrm{i}y,15,10^{15})$
for $\rho_1 = 15 \rightarrow \rho_2 = 10^{15}$.}
\end{figure}

Let us describe how we numerically approximate 
$\widehat{\phi}_\ell(\sigma,\rho_1,\rho_2)$ as a pole
sum $\widehat{\xi}_\ell^E(\sigma,\rho_1,\rho_2)$. 
A simple version of the procedure is to follow the steps listed 
in Algorithm \ref{tab:AGHcomp}, replacing step 3 with evaluations of 
the profiles 
$\mathrm{Re}\widehat{\phi}_\ell(\mathrm{i}y_j,\rho_1,\rho_2)$ and
$\mathrm{Im}\widehat{\phi}_\ell(\mathrm{i}y_j,\rho_1,\rho_2)$.
To generate these profiles, we use \eqref{eq:BHteleportK}, and
for each $y_j$ evaluation point perform the $\eta$ integration 
using a composite Gauss-Kronrod rule. 
Assuming $[A,B]$ is one
subinterval of $[\rho_1,\rho_2]$ in the composite 
rule, the details are as follows. Let $\eta = AB/[(B-A)q + A]$ for
$q \in [0,1]$, so that
\begin{equation}
\int_A^B \frac{\widehat{\omega}_\ell(\mathrm{i}y_j,\eta)}{\eta}
d\eta 
\simeq \sum_{p=1}^{N_\mathrm{GK}} \frac{\widehat{\omega}_\ell
\big(\mathrm{i}y_j,AB/[(B-A)q_p + A]\big)}{q_p + A/(B-A)}w_p,
\end{equation}
where $(q_p,w_p)_{p=1}^{N_\mathrm{GK}}$ are the 15 nodes and
weights for the Gauss-Kronrod rule relative to $[0,1]$, and by
the above equation we mean the same rule is separately applied 
to the real and imaginary parts of 
$\widehat{\omega}_\ell(\mathrm{i}y_j,\eta)$. This
numerical integration is accurate, because, in fact, for
each grid-point $y_j$ integration of both
Re$\widehat{\omega}_\ell(\mathrm{i}y_j,\eta)$ and
Im$\widehat{\omega}_\ell(\mathrm{i}y_j,\eta)$ always
involves terms of the same sign. With $N_C$ denoting the number 
composite subintervals, profiles for the RBC kernels on the
$y$-grid must be computed $N_\mathrm{GK}\cdot N_C$ times.
If $\rho_2 = \infty$ the last composite interval might
be handled through a semi-infinite quadrature.
Approximation of $\widehat{\theta}_\ell(\sigma,\rho)$
follows a procedure similar (although more complicated) to
the one outlined above for 
$\widehat{\phi}_\ell(\sigma,\rho_1,\rho_2)$.

Unfortunately, this simple procedure becomes too costly 
when $\rho_2 \gg \rho_1$. The problem is two-fold. First,
$N_C$ must be chosen large. Second, and more serious, the 
approximation window $[-y_\mathrm{max},y_\mathrm{max}]$ 
is fixed by the profiles for $\rho_1$ (the ``widest" profiles 
in the integration). However, since, as seen from 
Fig.~\ref{fig:RWell2_shrink}, the RBC profiles shrink as $\rho$ 
increases, the $y$-grid needs many adaptive levels to resolve 
the contribution to the $\eta$-integration from the profiles at
and near $\rho_2$. The $y$-grid then must have a large number 
of points, on which $N_\mathrm{GK}\cdot N_C$ complex function
evaluations are made (with each such evaluation costing 
$\simeq 10^7$ floating point operations). To bypass this issue, 
we follow another rather complicated procedure, whereby the 
interval $[\rho_1,\rho_2]$ is first broken up into $\mathcal{N}$ 
chunks $[\rho_\alpha,\rho_{\alpha+1}]$ which are typically 
decades like $[10^\alpha,10^{\alpha+1}]$ if $\rho_2$ is very 
large. Each chunk has its own approximation window 
$[-y^\alpha_\mathrm{max},y^\alpha_\mathrm{max}]$ and 
(relatively small) $y^\alpha$-grid, and we choose these 
to conform with how shrunken the profiles are for
$\rho \simeq \rho_\alpha$. Next, using the relatively simple 
procedure described in the last paragraph, for each of the 
$\mathcal{N}$ chunks we construct a compressed kernel (table) 
which approximates
$\widehat{\phi}_\ell(\mathrm{i}y,\rho_\alpha,\rho_{\alpha+1})$ as a 
sum of poles 
$\widehat{\xi}_\ell^E(\mathrm{i}y,\rho_\alpha,\rho_{\alpha+1})$.
The last step is to generate profiles (for the compression algorithm) 
on a large $y$-grid associated with a wide approximation window 
and sufficient resolution near the origin. However, these 
evaluations are now done via combination of all $\mathcal{N}$ 
tables. Therefore, they are drastically faster, 
since they are carried out through auxiliary evaluations 
made with the $\mathcal{N}$ distinct pole sums (rather than 
ODE integration). Finally, we note that the physical 
teleportation kernel used in a numerical simulation is
\begin{equation}
\Xi^E_\ell(t,r_1,r_2) = 
\frac{1}{2M}\xi^E_\ell(t/(2M),r_1/(2M),r_2/(2M)),\qquad
\end{equation}
where $\xi_\ell^E(\tau,\rho_1,\rho_2)$ is the
inverse Laplace transform (with respect to $\sigma$) of
$\widehat{\xi}_\ell^E(\sigma,\rho_1,\rho_2)$.

\subsubsection{Efficiency and storage} \label{sec:EffBH}

As either $\ell$ or $\varepsilon^{-1}$ becomes large the scaling 
observed in \cite{Lau1} for compressed, blackhole, RBC kernels
appears similar to the flatspace result \eqref{eq:AGHdscaling}
described in Sec.~\ref{sec:Eff}.
However, we stress that these are empirical 
observations, and there is no corresponding proof of 
\eqref{eq:AGHdscaling} or a similar result in the blackhole case.
Nevertheless, provided that the number of approximating 
exponentials in the blackhole case indeed grows sublinearly with 
both $\ell$ and the inverse $1/\varepsilon$ of the relative 
approximation error [cf.~Eq.~\eqref{eq:superr}], our implementation 
of RBC satisfies the same efficient scalings \eqref{eq:RBCscalings} 
established for flatspace compressed kernels. 

\begin{figure}
\includegraphics[clip=true,width=14.0cm]{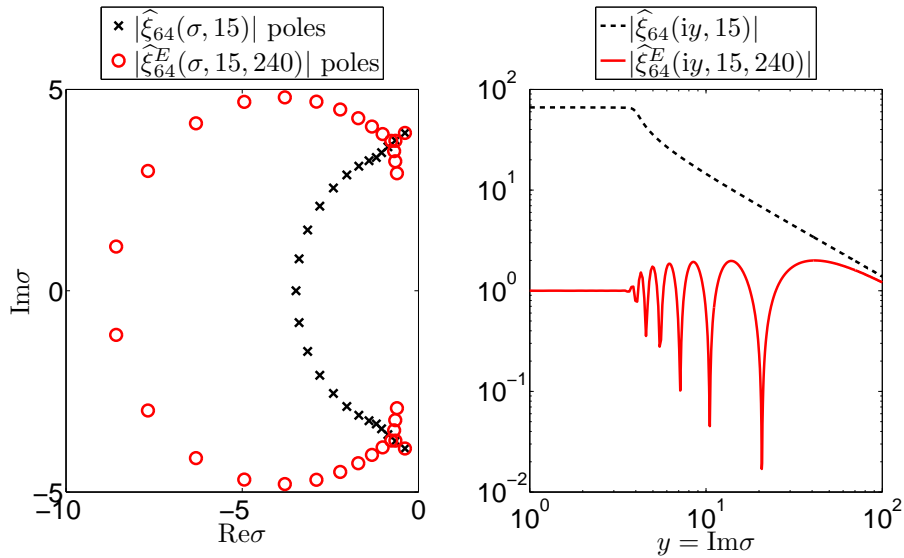}
\caption{{\sc Compressed RBC and teleportation kernels for $\ell = 64$.}
The right pane plots the modulus of the kernels for
positive $y$. For the RBC kernel (dotted line) this corresponds to the
modulus of the combined profiles shown in Fig.~\ref{fig:RWell64}, but
here with only half of the domain for the independent variable $y$. 
\label{fig:RBCandEXTell64kernComparison}}
\end{figure}

One might similarly ask whether or not our implementation of AWE 
for blackholes satisfies these scalings; we do not have
an answer for this question, but here consider kernels from 
the $\ell= 64$ teleportation experiment considered later in Section 
\ref{subsec:PulseT}. Let us focus on the compressed (frequency domain)
kernels $\widehat{\xi}{}_{64}(\sigma,15)$ and 
$\widehat{\xi}{}_{64}(\sigma,15,240)$, 
respectively for RBC at $\rho_b = 15$ and teleportation from 
$\rho_1 = 15$ to $\rho_2 = 240$. Figure \ref{fig:RWell64} has 
already depicted the profiles from which the 25-pole approximation 
$\widehat{\xi}{}_{64}(\sigma,15)$ is constructed. Notice that the
approximation $\widehat{\xi}{}_{64}(\sigma,15,240)$ has 32 poles,
whereas the corresponding exact {\em flatspace} teleportation kernel
would have 64 poles. Similar reduction for the $\ell = 64$ 
occurred for compressed flatspace kernels considered earlier.
As depicted in  Fig.~\ref{fig:RBCandEXTell64kernComparison} and
similar to the situation encountered in 
Fig.~\ref{fig:PolesFlatspaceRBCandTLPell64}, 
the pole locations for $\widehat{\xi}{}_{64}(\sigma,15)$ and 
$\widehat{\xi}{}_{64}(\sigma,15,240)$ are different.
Finally, we remark that we have tried to enforce the condition that 
the teleportation kernel has the same pole {\em locations}
as the RBC kernel [cf.~the discussion
around Eq.~\eqref{eq:WEandRBCpoles}]; however, we are then
unable to achieve a compressed kernel with any accuracy whatsoever.

The previous paragraph has considered the large-$\ell$, 
small-$\varepsilon$ limits. However, in this paper we mostly consider
$\varepsilon$ fixed (typically machine precision) and
small $\ell$, in which case, as we have seen, $d > \ell$. We remark
that this situation is similar to the case of ordinary wave propagation 
on $2+1$ flat spacetime. In that setting the low-$n$ (Fourier index) 
``circle kernels" are also expensive to evaluate, with
scalings similar to \eqref{eq:AGHdscaling}
only exhibited in the large $n$ limit \cite{AGH}. 

While the previous discussion has presented scalings for various limits, 
in practice our implementation of RBC/teleportation
for $\ell = 2,3,$ and $64$ 
amounts to adding on the order of 20 to 30 points to the spatial domain,
and modest increase in work and storage costs for the
numerical experiments considered here. 

\section{Numerical Experiments}\label{sec:Experiments}

To carry out numerical simulations, we have used both 
the nodal Legendre discontinuous Galerkin method described
in Ref.~\cite{FHL} (further details of this method will 
not be given here) and a nodal Chebyshev method. Both
methods feature multiple subdomains and upwinding.

\subsection{Pulse teleportation}\label{subsec:PulseT}
First consider the $\ell = 2$ Regge-Wheeler equation
with the same initial data \eqref{eq:GaussianData} given in 
Subsection \ref{subsec:QNRTails}. Using our multidomain nodal 
Chebyshev method, we perform five separate evolutions on 
domains with outer boundaries taken as the $b$ values 
corresponding to $r_b = 30M$, $60M$, $120M$, $240M$, and $480M$. 
We have respectively used 32, 37, 45, 62, and 95 subintervals 
of uniform size, and in each case with 32 Chebyshev-Lobatto 
points per subinterval. Therefore, the spatial 
resolution for each evolution is comparable to the others. 
Evolutions are performed by the classical 
4-stage explicit Runge Kutta method with timestep 
$\Delta t \simeq M(2.6794\times 10^{-4})$. For each 
evolution the inner boundary is $a = -200M$, and therefore the 
Sommerfeld boundary condition $-(\Pi+\Phi)= 0$ at the inner 
boundary is essentially exact. For all choices of outer boundary 
$b$ we adopt the Laplace convolution RBC \eqref{eq:LaplaceConvApprox}. 
Tables for $r_b = 30M$, $60M$, $120M$, $240M$, and $480M$ respectively 
have 19, 19, 19, 18, and 17 poles, with each table computed in 
quadruple precision to satisfy the tolerance $\varepsilon = 10^{-15}$. 
These tables are available at \cite{Kernel_web1}.
\begin{figure}
\includegraphics[clip=true,width=14.0cm]{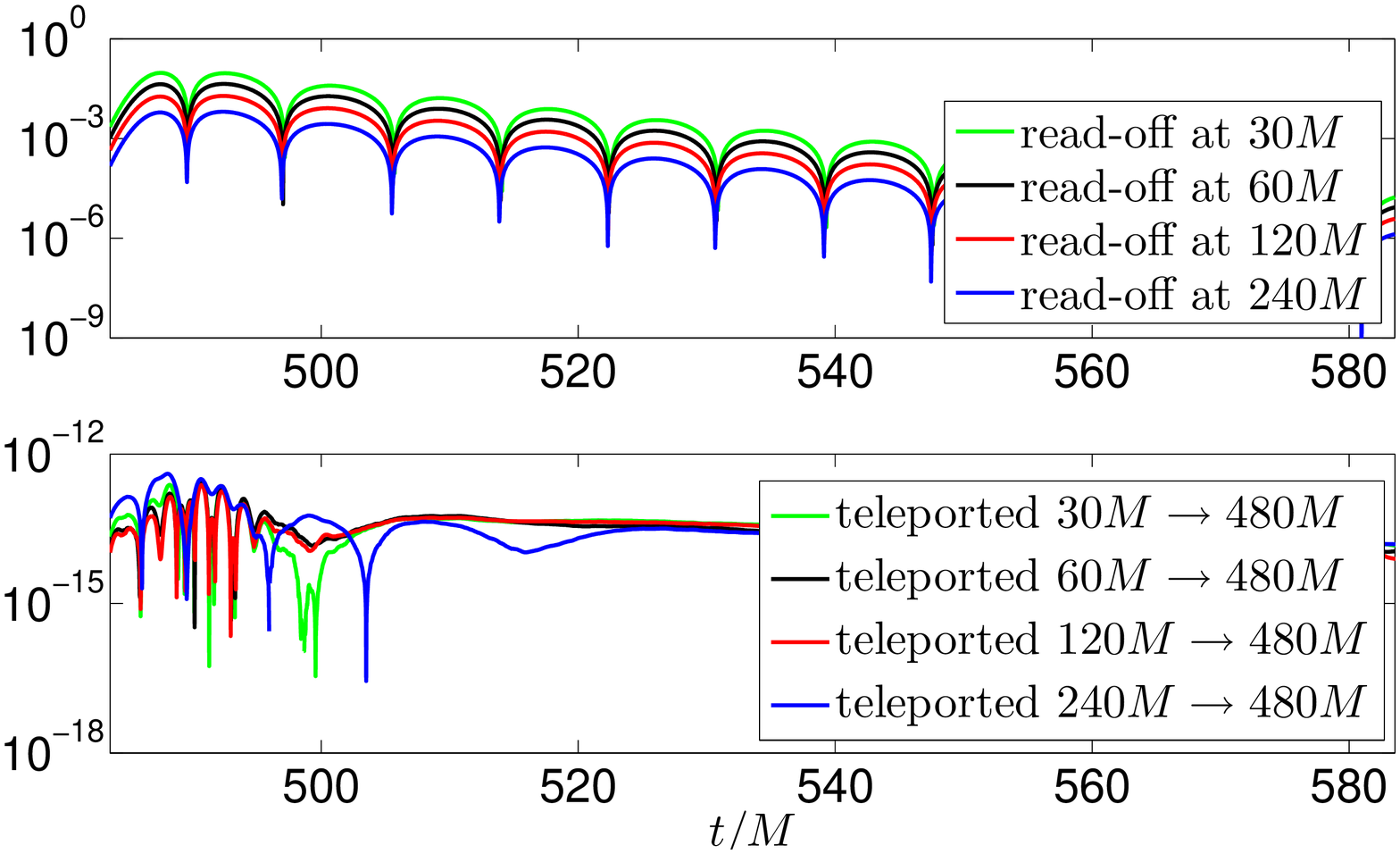}
\caption{{\sc Errors in read-off and teleported $\ell = 2$ waveforms
relative to read-off waveform at $r_2 = 480M$.}
\label{fig:PulseComparison2}
}
\end{figure}

In all cases the field $\Psi(t,r_b)$ is recorded as a 
time series at the boundary $b$, and in all cases but 
the last ($b$ corresponding to $r_b = 480M$)
we ``teleport" the field from $r_1 = r_b$ to $r_2 = 480M$.
Each approximate teleportation kernel 
$\Xi^E_2 \simeq \Phi_2$ features the same pole locations 
as the corresponding approximate RBC kernel 
$\widehat{\Xi}\simeq \Omega_2$. For the last $r_b = 480M$
simulation we simply record the field at the boundary, with 
this record then serving as a reference time series.
We account for time delays by starting all recorded times 
series (whether read off or teleported) at time $b - 6M$. 

The top panel in Figure \ref{fig:PulseComparison2} plots the 
errors in the waveforms recorded at the different $b$ 
boundaries as compared to the reference $r_b = 480M$ waveform; 
as expected the systematic errors are large.
The bottom panel plots the errors in the 
($r_1 = r_b$ to $r_2 = 480M$) teleported 
time series relative to the reference time series.
With the reference time series viewed as the ``asymptotic signal", 
this ``AWE" clearly yields 10 or more digits of 
accuracy relative to simple read-off. We have found similar results 
using other ``pulses" based on polynomial, Lorentzian, 
and trigonometric profiles (in all case with the initial 
data initially supported away from the boundary, either 
exactly or to machine precision).
\begin{figure}
\includegraphics[clip=true,width=14.0cm]{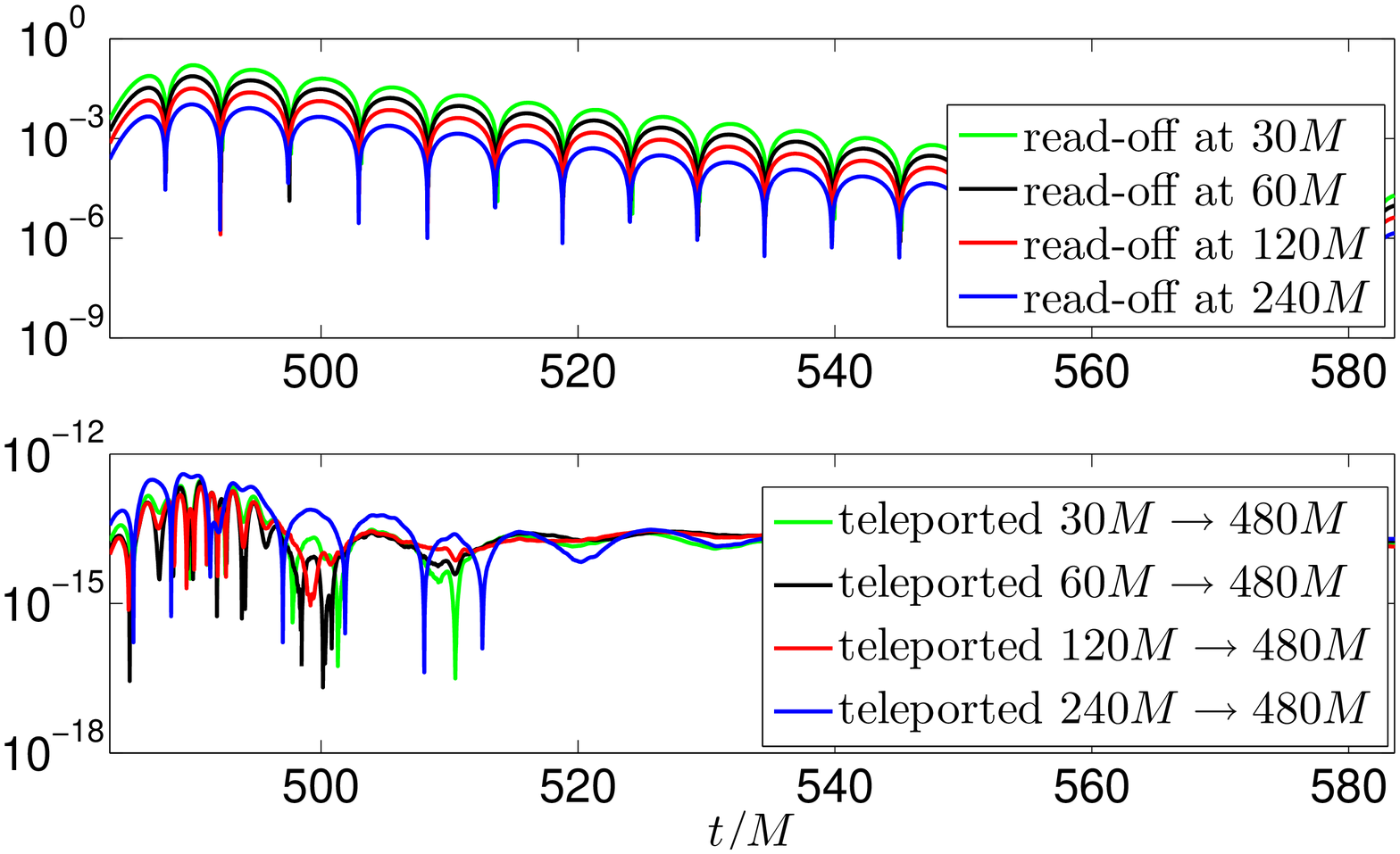}
\caption{{\sc Errors in read-off and teleported $\ell = 3$ waveforms
relative to read-off waveform at $r_2 = 480M$.}
\label{fig:PulseComparison3}
}
\end{figure}
\begin{figure}
\includegraphics[clip=true,width=14.0cm]{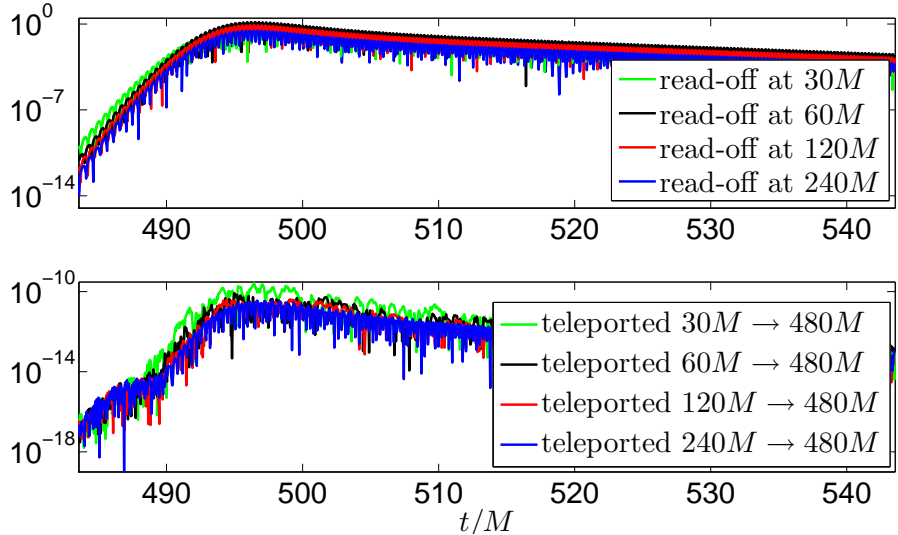}
\caption{{\sc Errors in read-off and teleported $\ell = 64$ waveforms
relative to read-off waveform at $r_2 = 480M$.}
\label{fig:PulseComparison64}
}
\end{figure}

We repeat the experiment for two different $\ell$ values. First, for 
$\ell = 3$ we adopt the same initial data and experimental setup, 
except for the numerical tables which specify the RBC and 
teleportation kernels. For the $\ell = 3$ experiment the 
number of poles for the RBC tables is either 15 or 16, and the 
number of poles for teleportation tables ranges from 15 to 20. 
The results, shown in Fig.~\ref{fig:PulseComparison3},
are comparable to those for $\ell = 2$. Lastly, we repeat the 
experiment for $\ell = 64$. For such a high $\ell$ the evolutions 
are much more expensive, due to finer oscillations in both space and 
time. We now use 42 points per subdomain, with the number of subdomains 
typically increased by a factor of 2 or 3 relative to the numbers
given above for $\ell = 2$. Moreover, we adopt the timestep 
$\Delta t = M(4.0461 \times 10^{-5})$ and inner boundary $a = -80M$
which is casually disconnected from each waveform read-off/teleportation 
at the outer boundary. For $\ell = 64$ 
our RBC tables have between 23 and 25 poles, and our teleportation 
tables between 30 and 32 poles. While these tables are large, 
note that even the corresponding {\em exact} flatspace RBC 
kernels would have 64 poles. Hence, in this experiment the savings 
afforded by kernel compression is already evident. Results are 
depicted in Fig.~\ref{fig:PulseComparison64}.

\subsection{Luminosities from extreme-mass-ratio binaries}
\label{subsec:luminosities}
An extreme mass ratio binary (EMRB) is a system comprised of 
a small mass-$m_p$ compact object (the ``particle") 
orbiting a much larger mass-$M$ blackhole, where the mass 
ratio $m_p/M \ll 1$. EMRB systems 
are expected to emit gravitational radiation in a low frequency 
band ($10^{-5}$ to $10^{-1}$ {\tt Hz}), and therefore offer the 
promise of detection by a space-based 
gravitational wave observatory like the earlier proposed LISA
project \cite{ScottHughes_LISA}. Located within the solar system,
such an observatory would be well approximated as positioned at
$\scri$ relative to expected sources.

A standard method for studying EMRBs uses the perturbation 
theory of Schwarzschild blackholes in an approximation which 
treats the particle as a point-like Dirac delta function. 
The particle follows a timelike 
geodesic in the background Schwarzschild spacetime and is 
responsible for generating small metric perturbations which 
radiate away 
(see Refs.~\cite{Martel_CovariantPert,SopuertaLaguna,Sarbach:2001qq}
for modern accounts of the subject).
Here we note that the axial metric perturbations 
for each $(\ell,m)$ mode may be combined to form a gauge 
invariant scalar quantity $\Psi_{\ell m}^{\mathrm{CPM}}$ which 
obeys Eq.~\eqref{eq:RWZeqn} with the Regge-Wheeler 
potential $V^\mathrm{RW}(r)$ 
and a distributional forcing term 
$S_{\ell m}^{\mathrm{CPM}}(t,r)$.\footnote{
\label{foot:RWCPM}
We use the Cunningham-Price-Moncrief (CPM) master function 
\cite{CPM:paper} which yields formulas for the axial sector which are on the 
same footing as those for the polar sector.}
Likewise, the polar metric perturbations for each $(\ell,m)$ 
mode may be combined to form a gauge invariant scalar 
quantity $\Psi_{\ell m}^{\mathrm{Z}}$ which obeys
Eq.~\eqref{eq:RWZeqn} with
the Zerilli potential $V^\mathrm{Z}(r)$ and a 
distributional forcing term $S_{\ell m}^{\mathrm{Z}}(t,r)$. 
Both $S^{\mathrm{Z}}_{\ell m}$ and $S^{\mathrm{CPM}}_{\ell m}$
are built from linear combinations of an $(\ell,m)$ mode
decomposition of the stress-energy tensor and, as such, 
depend on the particle trajectory 
$(r_p(t),\pi/2,\phi_p(t))$ in the equatorial plane. 
Bounded and stable orbits are characterized 
by an eccentricity constant $e$ and a semi-latus rectum 
constant $p$. Upon specification of $(e,p)$, the resulting 
trajectory is found by integrating the relevant system of 
ODEs given in Eq.~(5) of Ref.~\cite{FHL}. Appendix C of 
\cite{FHL} gives exact expressions
for $S_{\ell m}^{\mathrm{CPM,Z}}$ 
(see also \cite{SopuertaLaguna,Hopper:2010uv}).

With both $S_{\ell m}^{\mathrm{CPM,Z}}$ specified, 
we numerically solve for $\Psi_{\ell m}^{\mathrm{CPM,Z}}$, 
starting with trivial initial data and smoothly turning on 
the source term $S_{\ell m}(t,r)$ over 
a timescale $\tau \simeq 150M$ to $450M$
to prohibit static {\em Jost junk} solutions which may appear 
in some formulations when using inconsistent initial data
\cite{Field2010,PhysRevD.83.061503}. Respectively,
Sommerfeld and Laplace convolution RBC are enforced at the left and right 
physical boundaries (cf.~Sec.~\ref{subsec:simpleRBC}). The computational domain 
is given by the interval 
$[-400M,b]$, where the tortoise coordinate value $b$ 
corresponds to $r_b = 60M$ in Schwarzschild radius. 
Notice that as an approximation to the asymptotic 
signal at $\scri$, the waveform read-off at $r_b$ will 
have an $O(r_b^{-1})$ systematic error, suggesting relative errors 
greater than one percent for $r_b = 60M$.
For our simulations, we have chosen $16$ and $3$
subdomains to the left and right of the delta function 
respectively and represent the numerical solution by 
an order-$40$ or order-$46$ polynomial on each subdomain. 
The distributional source terms 
$S_{\ell m}^{\mathrm{CPM,Z}}$ determine jump conditions in 
the fields $\Psi_{\ell m}^{\mathrm{CPM,Z}}$ at the particle 
location which we impose as junction conditions between
subintervals ~\cite{FHL}, with the motion of the particle 
incorporated through a time-dependent coordinate transformation.
Our particular choice of $r_b = 60M$ ensures that the particle
does not come too close to the outer computational boundary
which might lead to over stretching of the coordinates.
Temporal integration is carried out with a classical 
fourth-order Runge-Kutta method with timestep 
$\Delta t = M(5 \times 10^{-3})$.

Computation of the luminosities for a particular orbit of the perturbing 
particle is a standard benchmark test. For each $(\ell, m)$
mode the energy and angular momentum luminosities at $\scri$
(denoted by $\infty$) and at the event horizon $r = 2M$ (denoted by $H$) 
are given by
\begin{subequations} \label{eqn:flux_mode}
\begin{align}
\dot{E}^{\infty,H}_{\ell m} & = \frac{1}{64 \pi}  
\frac{\left(\ell + 2\right)!}{\left(\ell - 2\right)!} 
\left< | \dot{\Psi}_{\ell m}^\mathrm{Z}|^2 
     + |\dot{\Psi}_{\ell m}^\mathrm{CPM}|^2 \right>, \\
\dot{L}^{\infty,H}_{\ell m} & = \frac{\mathrm{i}m}{64 \pi} 
\frac{\left(\ell + 2\right)!}{\left(\ell - 2\right)!} 
\left< \dot{\Psi}^\mathrm{Z}_{\ell m} \bar{\Psi}^\mathrm{Z}_{\ell m} 
     + \dot{\Psi}_{\ell m}^\mathrm{CPM} \bar{\Psi}^\mathrm{CPM}_{\ell m} \right> ,
\end{align}
\end{subequations}
where the average $\left< A(t) \right>$ of a time
series $A(t)$ is computed as
\begin{equation} \label{eqnAveragedQuant}
\langle A \rangle \equiv \frac{1}{T_2-T_1}\int_{T_1}^{T_2} A(t) dt,
\qquad 
T_2 - T_1 = T_r.
\end{equation}
Here $T_r$ is the period of radial oscillation for the particle orbit.

Before presenting our numerical results, we remark on the 
potential sources of error in an evaluated asymptotic waveform. 
At $b$ we record both $\Psi_{\ell m}^{\mathrm{CPM,Z}}(t,b)$ 
and their first temporal derivatives as a time-series. With 
the numerical setup described above, the relative 
pointwise error associated with these read-off waveforms 
is better than $10^{-10}$. An 
additional source of (systematic) error is due to 
trivial initial data which, both incorrect and inconsistent,
is known to generate spurious junk. At a finite and fixed 
radial location, spurious junk radiation propagates away (the 
potential for {\em static} Jost junk is discussed 
in Ref.~\cite{Field2010}), although due to backscattering 
a ``junk error tail" may develop which decays more slowly. Tail fields 
are expected to fall off like $t^{-4}$ at $\scri$, and $t^{-7}$ 
at a fixed (much smaller) radial value. Evidently, the situation 
is worse at $\scri$ where junk error tails decay more slowly. 
Additionally, we often need to average luminosity quantities over 
long periods of time. Taken 
together, these facts conspire to make the temporal average 
of a $\scri$ waveform especially prone to contamination by junk error 
tails, even at late times and especially when high accuracy is desired. 
Unfortunately, simply waiting for junk errors to die out may
not be practical, because ODE and PDE numerical integrators typically 
introduce numerical errors which grow linearly with time.
While convolution with an approximate teleportation 
kernel will introduce additional error, 
we believe that the dominant errors in our 
asymptotic waveforms stem from numerical method error and spurious junk.
\begin{table*}[h!]
{\scriptsize
\begin{tabular}{| cc || ll | ll | ll | ll |}
\hline
$m$ & Alg. 
& \multicolumn{2}{c|}{$\dot{E}^\infty_{2 m}$} 
& \multicolumn{2}{c|}{$\dot{E}^H_{2 m}$} 
& \multicolumn{2}{c|}{$\dot{L}^\infty_{2 m}$} 
& \multicolumn{2}{c|}{$\dot{L}^H_{2 m}$}   \\
\hline
\hline
0 	 & 	  FR 	 
& 	  1.27486196317 	 & 	 $\times 10^{-8}$ 	  
& 	  1.66171571270 	 & 	 $\times 10^{-8}$ 	  
&    \multicolumn{2}{c|}{0}
&    \multicolumn{2}{c|}{0} \\
 	 & 	 AWE 	 
& 	  1.27486196187		 & 	 $\times 10^{-8}$ 	  
& 	  1.66171571269		 & 	 $\times 10^{-8}$ 	  
&   \multicolumn{2}{c|}{0} 	
&   \multicolumn{2}{c|}{0}  \\
\hline
1 	 & 	 FR 	 
& 	  1.15338054092          & 	 $\times 10^{-6}$
& 	  3.08063328605          & 	 $\times 10^{-7}$
& 	  1.44066000650          & 	 $\times 10^{-5}$
& 	  2.77518962557          & 	 $\times 10^{-6}$\\
 	 & 	 AWE 	 
& 	  1.15338054091          & 	 $\times 10^{-6}$
& 	  3.08063328606          & 	 $\times 10^{-7}$
& 	  1.44066000619          & 	 $\times 10^{-5}$
& 	  2.77518962558          & 	 $\times 10^{-6}$\\
\hline
2 	 & 	 FR 	 
& 	  1.55967717209          & 	 $\times 10^{-4}$
& 	  1.84497995136          & 	 $\times 10^{-6}$
& 	  2.07778922470          & 	 $\times 10^{-3}$
& 	  1.85014840343          & 	 $\times 10^{-5}$\\
 	 & 	 AWE 	 
& 	  1.55967717211          & 	 $\times 10^{-4}$
& 	  1.84497995135          & 	 $\times 10^{-6}$
& 	  2.07778922439          & 	 $\times 10^{-3}$
& 	  1.85014840342          & 	 $\times 10^{-5}$ \\
\hline
\end{tabular}}
\caption{Mode-by-mode $\ell = 2$ luminosities for the
eccentric orbit described in the text. For a particle of 
mass $m_p$ these values should be scaled by $m_p^2$. 
The table compares our asymptotic-waveform 
evaluation (AWE) method with the accurate frequency domain (FR) luminosities. 
FR results refer to Table III of Ref.~\cite{Hopper:2010uv} 
and are quoted to a relative error of $10^{-12}$. 
For this experiment the outer boundary is $r_b = 60M$.
\label{highEFluxTable}
}
\end{table*}

Through $r_b \rightarrow r_{\infty} = 2M(1 \times 10^{15})$ 
teleportation, we approximately obtain the 
signals $\Psi_{\ell m}^{\mathrm{CPM,Z}}$ at $\scri$, and 
with these signals compute the energy and angular momentum 
luminosities (\ref{eqn:flux_mode}a,b).
The orbital parameters $(e,p) = (0.764124,8.75455)$ and 
initial location $(r_p, \phi_p) = ((pM)/(1+e\cos(\pi/2)),0)$
specify the particle's path.
As described above, we slowly turn-on the distributional source 
over a timescale of $\tau \simeq 150M$ to $450M$. A physically
meaningful luminosity measurement will not depend on our choice 
of $\tau$, and from this consideration we find that by $3000M$ 
the spurious junk's effect is minimal. Table~\ref{highEFluxTable} 
compares the $\ell=2$ luminosity measurements at $\scri$ with 
the accurate frequency-domain results reported in 
Ref.~\cite{Hopper:2010uv}. We match their 
stated accuracy to better than $9$ digits.

As our final experiment we consider a circular orbit 
specified by the orbital parameters $(e,p) = (0,10)$. 
Circular-orbit luminosity measurements are time-independent, 
thereby allowing us to (i) better understand the influence of 
junk error tails on $\scri$ waveforms and (ii) estimate 
errors due to our AWE procedure in a clean setting.
With the same numerical parameters used for our eccentric 
orbit simulations, we compute $\ell=2$ 
luminosities at $\scri$ and compare them to accurate
frequency-domain results generated with the code 
described in Ref.~\cite{Hopper:2010uv}. By $T=6000M$ 
our results agree with the frequency-domain
results to within a relative difference of less than $10^{-12}$. 
Furthermore, we find the same level of agreement when the 
outer boundary is moved inward to $r_b = 30M$, 
in which case we use the teleportation/RBC 
kernel tables given in Appendix~\ref{App:Tables} 
(with the longer teleportation table for AWE).

\begin{figure}
\includegraphics[clip=true,width=0.48\linewidth]{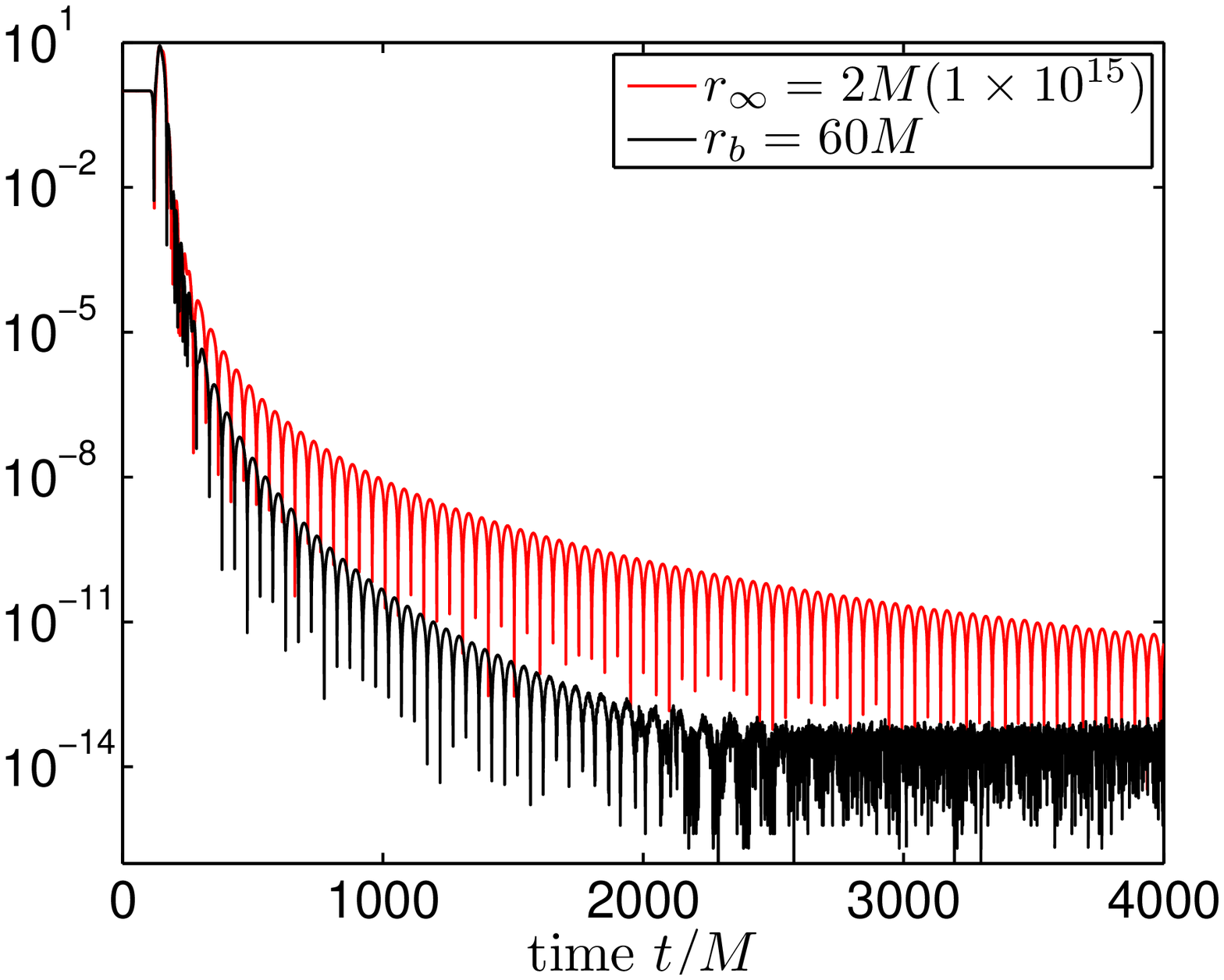}
\includegraphics[clip=true,width=0.48\linewidth]{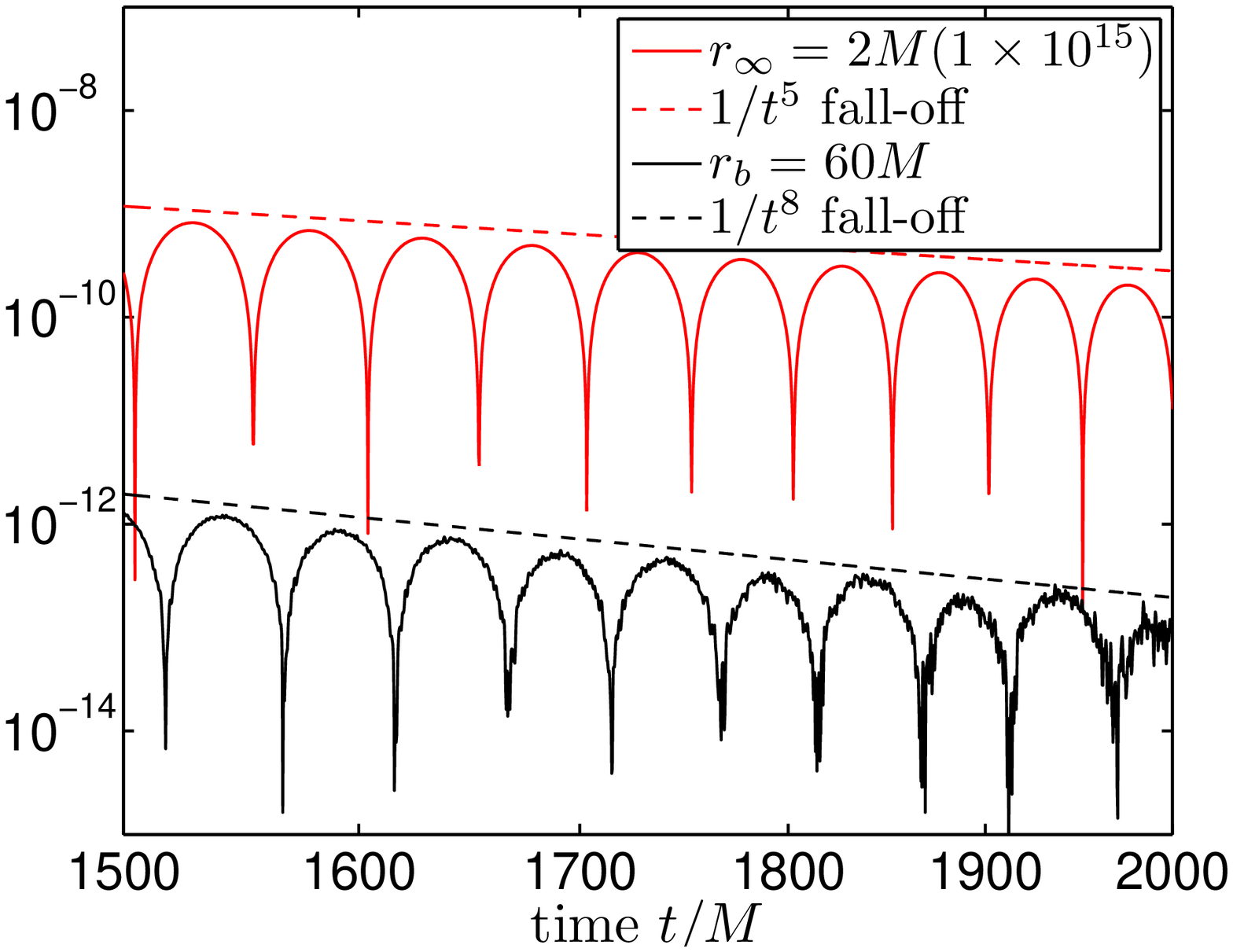}
\caption{{\sc Energy luminosity errors due to spurious junk radiation.}
Both panels show
$|\dot{E}^{\infty}_{22}(t) - \dot{E}^{\infty}_{2 2,\mathrm{FR}}|
/\dot{E}^{\infty}_{22,\mathrm{FR}}$ (red line, $r_{\infty}$) and
$|\dot{E}^{b}_{2 2}(t) - \dot{E}^{b}_{2 2}(6500M)|
/\dot{E}^{b}_{2 2}(6500M)$ (black line, $r_b$) as a time-series.
Comparing the black and red lines, we see that the energy 
luminosity error (due to spurious junk) is particularly persistent
at $\scri$. See the text for further explanation. 
\label{fig:Junktails}}
\end{figure}

As the final measurement time is taken earlier, 
the agreement becomes progressively worse due to 
spurious junk radiation. Indeed, the solid black line in 
Fig.~\ref{fig:Junktails} (left) plots the relative error 
$|\dot{E}^{\infty}_{22}(t) 
- \dot{E}^{\infty}_{22,\mathrm{FR}}|\big/
  \dot{E}^{\infty}_{22,\mathrm{FR}}$
as a time-series, where $\dot{E}^{\infty}_{2 2,\mathrm{FR}}$ is 
the frequency-domain value. For comparison we also compute 
a ``luminosity" quantity\footnote{At finite radial values, 
especially ones this small, $\dot{E}^{b}_{2 2}$ is certainly not 
the energy radiated by the system. However, this value is computable 
and, furthermore, is theoretically (although perhaps not 
numerically) constant for circular orbits.
Therefore, our intention here is to quantify the effect of junk error tails 
on its computation. Approximation of $\dot{E}^{\infty}_{2 2}$, 
perhaps by extrapolation, might rely on such measurements.}
$\dot{E}^{b}_{2 2}(t)$ from $\dot{\Psi}_{2 2}^\mathrm{Z}(t,b)$. 
The solid black line in Fig.~\ref{fig:Junktails} (left) 
shows the relative error 
$|\dot{E}^{b}_{2 2}(t) 
- \dot{E}^{b}_{2 2}(6500M)|\big/
  \dot{E}^{b}_{2 2}(6500M)$, 
where $\dot{E}^{b}_{2 2}(6500M)$ is a late-time
computation less contaminated
by spurious junk. Comparing the black 
and red lines, we see that the junk error tails 
at $\scri$ persist longer than those at the outer boundary $b$. This
observation suggests that spurious junk radiation is 
a stubborn problem for high accuracy studies. 
The right panel of Fig.~\ref{fig:Junktails}
indicates that the energy luminosity errors (due to spurious junk) 
respectively decay as $t^{-8}$ and $t^{-5}$ 
for a fixed radial value and $\scri$. 
If we view $\dot{\Psi}_\mathrm{exact}$ 
as either $\dot{\Psi}_{22}(6500,b)$ or $\dot{\Psi}_{22}^\infty$, then 
these rates are consistent with the expected decay rate 
for {\em field} error tails $\Psi_\text{junk tail}$ 
and the relationship for a numerically computed 
energy luminosity $\dot{E} \propto | \dot{\Psi}_\mathrm{exact}
+ \dot{\Psi}_\text{junk tail}|^2 \simeq
|\dot{\Psi}_\mathrm{exact}|^2 + 2\mathrm{Re}\big(\dot{\Psi}_\mathrm{exact}
\dot{\Psi}_\text{junk tail}\big)$.
\begin{table*}[h!]
{\scriptsize
\begin{tabular}{| cc || ll | ll | ll | ll |}
\hline
$m$ & Alg. 
& \multicolumn{2}{c|}{$\dot{E}^\infty_{2 m}$} 
& \multicolumn{2}{c|}{$\dot{E}^H_{2 m}$} 
& \multicolumn{2}{c|}{$\dot{L}^\infty_{2 m}$} 
& \multicolumn{2}{c|}{$\dot{L}^H_{2 m}$}   \\
\hline
\hline
1 	 & 	 FR 	 
& 	  1.93160935116          & 	 $\times 10^{-7}$
& 	  1.22691683145          & 	 $\times 10^{-9}$
& 	  6.10828509933          & 	 $\times 10^{-6}$
& 	  3.87985168700          & 	 $\times 10^{-8}$\\
 	 & 	 AWE 	 
& 	  1.93160935114          & 	 $\times 10^{-7}$
& 	  1.22691683145          & 	 $\times 10^{-9}$
& 	  6.10828509953          & 	 $\times 10^{-6}$
& 	  3.87985168700          & 	 $\times 10^{-8}$\\
\hline
2 	 & 	 FR 	 
& 	  5.36879547910          & 	 $\times 10^{-5}$
& 	  1.13082774691          & 	 $\times 10^{-8}$
& 	  1.69776220056          & 	 $\times 10^{-3}$
& 	  3.57599132155          & 	 $\times 10^{-7}$\\
 	 & 	 AWE 	 
& 	  5.36879547910          & 	 $\times 10^{-5}$
& 	  1.13082774691          & 	 $\times 10^{-8}$
& 	  1.69776220057          & 	 $\times 10^{-3}$
& 	  3.57599132154          & 	 $\times 10^{-7}$ \\
\hline
\end{tabular}}
\caption{Mode-by-mode $\ell = 2$ luminosities for
the circular orbit described in the text.
For a particle of mass $m_p$ these values should be scaled by $m_p^2$. 
The table compares our asymptotic-waveform evaluation
(AWE) results with frequency domain (FR) results computed by the code described 
in Ref.~\cite{Hopper:2010uv}. We are grateful to S.~Hopper for 
generating these previously unpublished FR luminosity values.
\label{highEFluxTable_Circ}
}
\end{table*}

\section{Conclusions} \label{sec:conclude}

In the context of the Regge-Wheeler/Zerilli (describing blackhole 
perturbations) and ordinary (describing acoustic phenomena) wave 
equations we have developed a procedure 
for obtaining the asymptotic far-field signal from a time-series 
recorded at a finite radial value located beyond the spatial 
compact support of the initial data. Furthermore, we have viewed 
asymptotic waveform evaluation as a limiting
case of signal teleportation between two 
finite radial values. For each of these wave equations our steps 
are to (i) write down the exact relationship for teleportation in 
the Laplace frequency domain, (ii) approximate this relationship 
along the inversion contour (of the inverse Laplace transform) by 
a sum of simple poles, and (iii) then represent, through inversion,
the asymptotic signal as a convolution [cf.~Eq.~\eqref{eq:XiEkernelConv}] 
of the solution with time-domain kernel [cf.~Eq.~\eqref{eq:XiEkernel}]
comprised of damped exponentials. A similar recipe might be 
used both to impose boundary conditions for and evaluate asymptotic 
waveforms from perturbations of a Kerr blackhole. The Teukolsky 
equation describing such perturbations is, like the cases treated 
in this paper, separable in the frequency domain.

Through pre-computed numerical tables specifying each exponential's 
strength and damping rate, we have demonstrated that accurate 
asymptotic waveform evaluation through
teleportation can be easily implemented. Our simulations 
based on these numerical tables correctly exhibit $t^{-4}$ as the 
asymptotic decay rate for $\ell=2$ tails. We have also performed 
generic-orbit, extreme-mass-ratio, binary simulations. From the solution 
recorded as close as $30M$ and $60M$, we compute far-field 
luminosities which agree with accurate frequency domain results to a 
relative error of better than $10^{-9}$ ($10^{-12}$ for circular 
orbits). Our studies indicate that spurious junk radiation is 
particularly problematic for $\scri$ computations, because far-field
luminosity errors (due to spurious junk) decay at a slow rate.
These results have been achieved without a compactification 
scheme to include $\scri$ in the computational domain. Instead, 
they have relied on a Laplace convolution \eqref{eq:XiEkernelConv}
which, decoupled from a numerical evolution, can also be carried 
out as a post-processing step on an existing time series. 
Finally, we have demonstrated effective signal teleportation between
two finite radial locations, for $\ell = 2,3,64$ and with relative
errors $\simeq 10^{-10}$. These demonstrations are a powerful test of 
our AWE/teleportation method's 
accuracy as well as a practical sanity check of its implementation.

As discussed in Sec.~\ref{sec:EffBH}, our Laplace-convolution RBC 
and AWE methods would seem efficient from work and storage 
standpoints. Lower $\ell$ kernels appear similar to the case of ordinary 
wave propagation on $2+1$ flat spacetime. In that setting the low-$n$ 
(Fourier index) ``circle kernels" are also expensive to evaluate to 
account for tail-like phenomena. In any case, for the $\ell =
2,3,64$ cases considered in this paper a kernel's overall computational
cost is roughly equivalent to to adding 20 to 30 points to 
the spatial domain. Moreover, it is non-intrusive, requiring no grid 
stretching or supplemental coordinate transformations, and 
may be carried out at any radial value beyond the spatial compact support
of the initial data and sources. Finally, in their
frequency-domain form, our kernels might be used to 
implement radiation boundary conditions and AWE
in {\em frequency-domain codes.} Here we envision that
the kernels would first undergo a ``Wick rotation" prior
to use.

While the results of this paper are encouraging, we believe that 
further careful study is merited. First, the task of computing 
teleportation/AWE tables is daunting for a number of reasons. 
The main one is cost. Here we refer to the {\em offline} cost in
generating a table, not the cost incurred by the user implementing 
AWE with such a table. As discussed in Sec.~\ref{subsec:BHWE}, 
generation of AWE kernels costs upwards of $10^{13}$ 
floating point operations. Moreover, since the cost is offline, 
with the resulting numerical table then ``good for all time", we believe 
the process should be carried out in quadruple precision in order to achieve 
$\varepsilon = 10^{-15}$ error tolerances.\footnote{Indeed, once such 
accurate tables have been constructed, smaller tables (say corresponding 
to $\varepsilon = 10^{-8}$) can be achieved by compressing the accurate 
table, i.e.~using the accurate table for fast evaluation of profiles that 
are again subject to AGH compression.} This further adds to the cost.
Due to the difficulties associated with the computation of AWE tables, 
we have not adequately isolated all sources of error in their construction. 
A systematic and optimized procedure for computing kernels would greatly 
reduce the offline costs. One possibility is an application specific 
quadrature rule. So far, we have employed the familiar Gauss-Kronrod rule, 
which is designed for high-order integration of polynomials. We might instead 
design a quadrature rule which is exact for the corresponding 
flatspace kernels~\cite{Antil:ROQ}.

We plan to construct a family of RBC and AWE 
tables for general use: 
Regge-Wheeler and Zerilli tables likely for $2 \leq \ell \leq 64$, 
boundary radii $\rho_b = 15,30,60,120$, 
and an evaluation radius $\rho_\infty = 1\times 10^{15}$ (or 
$\rho_\infty = \infty$ with a semi-infinite quadrature rule, 
see Sec.~\ref{subsec:BHWE}). All kernels used in this paper, as well as these
others, will be available at \cite{Kernel_web1}.

\section{Acknowledgements}
This work has been substantially supported by NSF grant No.~PHY
0855678 to the University of New Mexico; AGB and SRL gratefully
acknowledge this support. AGB's contributions to this work
were also supported by NSF Grant No.~0739417 as an Undergraduate 
Research Project supervised by SRL. SEF acknowledges support from 
the Joint Space Science Institute, 
NSF Grants No.~PHY 1208861 and No.~PHY 1005632 to the University of
Maryland, and NSF grant No.~PHY05-51164 to the University of 
California at Santa Barbara. SRL is also grateful for support from 
the Erwin Schr\"{o}dinger Institute for Mathematical Physics, 
Vienna, during the workshop {\em Dynamics of General Relativity 
Analytical and Numerical Approaches}, 4 July to 2 September, 2011.
SEF thanks the Kavli Institute for Theoretical Physics, University 
of California at Santa Barbara, where this work was completed, 
for its hospitality. We are grateful for comments 
from M.~Blair, P.~Ca\~nizares, M.~Cao, C.~Evans, T.~Hagstrom, S.~Hopper,
L.~Lehner, M.~Tiglio, B.~Whiting, and A.~Zengino\u{g}lu. We further
thank A.~Zengino\u{g}lu for comments on the manuscript, and SRL 
particularly thanks C.~Evans for discussions early on in the 
development of RBCs which also influenced the ideas presented here.

\appendix

\section{Error estimates} \label{sec:ErrorEst}
Suppose that we have an ``exact" kernel $B(t)$ and an 
associated Laplace convolution
\begin{equation}
(B*\Psi)(t) = \int_0^t B(t-t')\Psi(t') dt'.
\end{equation}
We then have the following result for the relative 
convolution error associated with using an 
{\em approximate kernel} $A(t)$ in place of $B(t)$:
\begin{equation}\label{eq:basic_estimate_rel}
\|A*\Psi-B*\Psi\|_{L_2(0,\infty)} \leq 
\sup_{s\in\mathrm{i}\mathbb{R}}
\frac{|\widehat{A}(s)-\widehat{B}(s)|}{|\widehat{B}(s)|}
\|B*\Psi\|_{L_2(0,\infty)},
\end{equation}
provided that $\widehat{B}(s)\neq 0$ holds for all 
$s \in\mathrm{i}\mathbb{R}$. If this condition fails, 
we have instead
\begin{equation}\label{eq:basic_estimate_abs}
\|A*\Psi-B*\Psi\|_{L_2(0,\infty)} \leq
(2\pi)^{-1/2}
\sup_{s\in\mathrm{i}\mathbb{R}}
|\widehat{A}(s)-\widehat{B}(s)|\cdot\|\Psi\|_{L_2(0,\infty)},
\end{equation}
Before discussing their consequences, let us verify 
\eqref{eq:basic_estimate_rel} and 
\eqref{eq:basic_estimate_abs}.

A Laplace convolution may be viewed as a Fourier 
convolution, that is
\begin{equation}
\int_0^t B(t-t')\Psi(t') dt' 
= \int_{-\infty}^\infty B(t-t')\Psi(t') dt',
\end{equation}
if we adopt that viewpoint that $B(t)$ and $\Psi(t)$ are 
{\em causal functions}, i.e.~$B(t) = B(t)\theta(t)$ and 
$\Psi(t) = \Psi(t)\theta(t)$, where $\theta(t)$ is the 
Heaviside step function. With this viewpoint, the Fourier 
transform of $\Psi(t)$, for example, is
\begin{equation}
\widetilde{\Psi}(\omega) = \frac{1}{\sqrt{2\pi}}
\int_0^\infty \exp(-\mathrm{i}\omega t) \Psi(t) dt,
\end{equation}
with the following formal relationship holding between 
the Fourier and Laplace transforms:\footnote{Although 
its neglection does not spoil the final estimate, the 
factor of $1/\sqrt{2\pi}$ was neglected on page 4156 
of Ref.~\cite{Lau2} (and on pages 23 and 24 of 
{\tt arXiv:gr-qc/0401001v3}).}
\begin{equation}
\widetilde{\Psi}(\omega) = 
\frac{1}{\sqrt{2\pi}} \widehat{\Psi}(\mathrm{i}\omega).
\end{equation}
To establish \eqref{eq:basic_estimate_rel}, we view 
$A*\Psi$ and $B*\Psi$ as Fourier convolutions of casual 
functions (each convolution is again a causal function), 
and with the Parseval and Fourier convolution theorems 
find that
\begin{align}
\begin{split}
\|A*\Psi-B*\Psi\|_{L_2(0,\infty)} 
& = \|A*\Psi-B*\Psi\|_{L_2(\mathbb{R})}\\
& = \|\widetilde{A}\widetilde{\Psi}
- \widetilde{B}\widetilde{\Psi}\|_{L_2(\mathbb{R})}\\
& \leq \sup_{\omega \in \mathbb{R}}
\left| \frac{\widetilde{A}(\omega) 
            -\widetilde{B}(\omega)}{
             \widetilde{B}(\omega)} 
\right|
\|\widetilde{B}\widetilde{\Psi}\|_{L_2(\mathbb{R})} \\
& = \sup_{s\in\mathrm{i}\mathbb{R}}
\left| \frac{\widehat{A}(s)
            -\widehat{B}(s)}{
             \widehat{B}(s)} 
\right|
\|\widetilde{B}\widetilde{\Psi}\|_{L_2(\mathbb{R})}.                    
\end{split}
\end{align}
Using the inverse Fourier transform on the final term
to work backwards (again with the Parseval and Fourier 
convolution theorems), we obtain \eqref{eq:basic_estimate_rel}. 
The alternative estimate \eqref{eq:basic_estimate_abs} follows 
by nearly the same calculation.

We view \eqref{eq:basic_estimate_rel} as an estimate for either
of the error quantities 
\begin{subequations}\label{eq:E_1_2}
\begin{align}
\mathcal{E}_1 & \equiv 
                \big\| \Omega_\ell(\cdot,r_b)*\Psi_\ell(\cdot,r_b) 
              - \Xi_\ell(\cdot,r_b)*\Psi_\ell(\cdot,r_b)\|_{L_2(0,\infty)}\\
\mathcal{E}_2 & \equiv
                \big\| \Phi_\ell(\cdot,r_b,r_\infty)*\Psi_\ell(\cdot,r_b) 
              - \Xi^E_\ell(\cdot,r_b,r_\infty)*\Psi_\ell(\cdot,r_b)
                \|_{L_2(0,\infty)},
\end{align}
\end{subequations}
and \eqref{eq:basic_estimate_abs} as an estimate for
\begin{align} \label{eq:E_3}
\mathcal{E}_3 & \equiv
                \big\|
                \Phi_\ell(\cdot,r_b,r_\infty)*\Psi_\ell(\cdot,r_b) 
              - \Phi_\ell(\cdot,r_b,\infty)*\Psi_\ell(\cdot,r_b)
                \|_{L_2(0,\infty)}.
\end{align}
Quantities (\ref{eq:E_1_2}a,b) measure the quality of our numerical 
approximations for RBC and teleportation kernels, 
while \eqref{eq:E_3} measures the quality of using the exact waveform at a
large (but finite) radius $r_\infty \neq \infty$ as an approximation 
for the asymptotic waveform at $\scri$. We only present details for 
(\ref{eq:E_1_2}a) and \eqref{eq:E_3}.

To use the estimate \eqref{eq:basic_estimate_rel} for (\ref{eq:E_1_2}a), 
let $B(t) = \Omega_\ell(t,r_b)$ and $A(t) = \Xi_\ell(t,r_b)$.
Using Algorithm \ref{alg:compressRBC} we approximate 
$\widehat{B}(\mathrm{i}y) = \widehat{\Omega}_\ell(\mathrm{i}y,r)$ along 
the axis of imaginary Laplace frequency, demanding that
\begin{equation}\label{eq:freqboundRBC}
\sup_{y\in\mathbb{R}}
\frac{|\widehat{\Omega}_\ell(\mathrm{i}y,r)
      -\widehat{\Xi}_\ell(\mathrm{i}y,r)|}{
      |\widehat{\Omega}_\ell(\mathrm{i}y,r)|}
< \varepsilon.
\end{equation}
Here $\widehat{\Xi}_\ell(\mathrm{i}y,r) = \widehat{A}(\mathrm{i}y)$ 
is a sum of simple poles specified by one of our numerical tables, 
and $\varepsilon$ is the desired tolerance. This formula is essentially 
\eqref{eq:superr} written earlier in dimensionless variables for the 
blackhole case. With the above identifications, Eqs.~\eqref{eq:freqboundRBC} 
and \eqref{eq:basic_estimate_rel} yield
\begin{equation}
\mathcal{E}_1 < \varepsilon
\big\|\Omega_\ell(\cdot,r_b)*\Psi_\ell(\cdot,r_b)\|_{L_2(0,\infty)}.
\end{equation}
Note that \eqref{eq:freqboundRBC} is an {\em a posteriori} bound;
it is verified in step 5 of Algorithm \ref{alg:compressRBC}.

To facilitate the analysis for \eqref{eq:E_3}, we use $r_1$ and $r_2$
in place of $r_b$ and $r_\infty$, with $B(t) = \Phi_\ell(t,r_1,\infty)$ 
and $A(t) = \Phi_\ell(t,r_1,r_2)$. 
Now referring to the absolute estimate 
\eqref{eq:basic_estimate_abs}, we must control the factor
\begin{equation}
\big|\widehat{A}(s)-\widehat{B}(s)\big|
= 
\left|
\frac{W(sr_2,2Ms)}{W(sr_1,2Ms)} - \frac{1}{W(sr_1,2Ms)}
\right|,
\end{equation}
here expressed for the blackhole case 
[cf.~Eq.~\eqref{eq:BHteleportK}]. Further analysis of the
blackhole case would presumably rely on the known asymptotic
expansions (see footnote \ref{fn:asymptotic_series}) for
$W(sr,2Ms) = W(\sigma\rho,\sigma)$, but would seem difficult.
Therefore, to proceed, we switch to the simpler flatspace case 
by setting $M=0$ and using the formal result $W(sr,0) = W(sr)$. 
The expression
\eqref{eq:ellmult_fd} for $W_\ell(sr)$ then gives
\begin{equation}
\big|
\widehat{A}(s)-\widehat{B}(s)\big|
= \left|
\frac{
\sum_{k=1}^\ell c_{\ell k}(sr_2)^{-k}
}{
\sum_{j=0}^\ell c_{\ell j}(sr_1)^{-j}
}
\right|
=
\left(\frac{r_1}{r_2}\right)^\ell
\left|
\frac{
\sum_{k=1}^\ell c_{\ell k}(sr_2)^{\ell-k}
}{
\sum_{j=0}^\ell c_{\ell j}(sr_1)^{\ell-j}
}
\right|.
\end{equation}
We now show that
\begin{equation}
\sup_{s\in\mathrm{i}\mathbb{R}}
\left(\frac{r_1}{r_2}\right)^\ell
\left|
\frac{
\sum_{k=1}^\ell c_{\ell k}(sr_2)^{\ell-k}
}{
\sum_{j=0}^\ell c_{\ell j}(sr_1)^{\ell-j}
}
\right| = O(1/r_2).
\end{equation}
To establish this claim, note that the denominator
of the expression inside the operation of complex modulus
is the Bessel polynomial with zeros $b_{\ell,k}$. Therefore,
we expand the expression as a sum of simple poles, 
thereby finding
\begin{equation}
\sup_{s\in\mathrm{i}\mathbb{R}}
\left(\frac{r_1}{r_2}\right)^\ell
\left|
\frac{
\sum_{k=1}^\ell c_{\ell k}(sr_2)^{\ell-k}
}{
\sum_{j=0}^\ell c_{\ell j}(sr_1)^{\ell-j}
}
\right| \leq
\sup_{s\in\mathrm{i}\mathbb{R}}
\sum_{j=1}^\ell\left|\frac{\mu_j(r_1,r_2)}{sr_1 - b_{\ell,j}}\right|
\leq
\sum_{j=1}^\ell\left|\frac{\mu_j(r_1,r_2)}{\mathrm{Re}b_{\ell,j}}\right|.
\end{equation}
The residue formula for a simple pole shows that 
each coefficient in the expansion obeys $\mu_j(r_1,r_2) = O(1/r_2)$,
establishing the claim. These calculations show that (returning to 
$r_b$ and $r_\infty$ for $r_1$ and $r_2$)
\begin{equation}
\mathcal{E}_3  = O(1/r_\infty) 
\|\Psi_\ell(\cdot,r_b)\big\|_{L_2(0,\infty)}.
\end{equation}
While we have not proved a similar formula for the blackhole
case, this formula (an {\em a priori} estimate) has motivated 
our choice $r_\infty = 2M (1\times 10^{15})$ for 
double precision arithmetic.

\section{Derivation of NRBC without Laplace transform}
\label{sec:NoLaplace}
This appendix derives the nonlocal nonreflecting boundary 
condition \eqref{eq:SommerfeldRes} without appealing to 
the Laplace transform. In order to elucidate the main ideas,
we choose to focus on the representative $\ell = 2$ case;
the derivation for generic $\ell$ is more cumbersome but
similar. The Sommerfeld residual for the solution 
(\ref{eq:ell2mult_td}) is 
\begin{equation}
\partial_t\Psi_2 + \partial_r\Psi_2 
= -\frac{3}{r^2}f^{(1)}(t-r) - \frac{6}{r^3}f(t-r),
\label{SommerRes}
\end{equation}
and we will show that this equation can be expressed as
\begin{align}
& \partial_t\Psi_2 + \partial_r\Psi_2
= 
\frac{1}{r}\int_0^t \Omega_2(t-t',r)\Psi_2(t',r)dt'
\label{eq:full_conv}\\
& -\exp\Big(-\frac{3t}{2r}\Big)\Big\{
\sin\Big(\frac{\sqrt{3}t}{2r}\Big)
\Big[\frac{\sqrt{3}}{r^2}f^{(1)}(-r)
\Big]
+ \cos\Big(\frac{\sqrt{3}t}{2r}\Big)
\Big[\frac{3}{r^2}f^{(1)}(-r)
     + \frac{6}{r^3}f(-r)
\Big]\Big\},
\nonumber
\end{align}
where the $\ell = 2$ time-domain kernel is 
[cf.~\eqref{eq:Omegahat2}]
\begin{equation}\label{eq:exactROBCFLTl2}
\Omega_2(t,r) = \frac{z_+}{r}
\exp\left(\frac{z_+}{r}t\right)
+ \frac{z_-}{r}
\exp\left(\frac{z_-}{r}t\right),
\quad
z_\pm = -\frac{3}{2}\pm \mathrm{i}\frac{\sqrt{3}}{2}.
\end{equation}
Let us postpone establishing \eqref{eq:full_conv}, and first consider
its consequences. If we assume \eqref{eq:IDassume} and evaluate 
\eqref{eq:full_conv} at $r = r_b$, then the last two terms on the 
righthand side vanish and we have the desired result
\begin{equation}
\big[\partial_t\Psi_2(t,r) + \partial_r \Psi_2(t,r)\big]\big|_{r=r_b}
= \frac{1}{r_b}\big[\Omega_2(\cdot,r_b) * \Psi_2(\cdot,r_b)\big](t).
\end{equation}
Indeed, from (i) the identity
\begin{equation}
f(t-r) = \frac{1}{6}r^2 (\partial_t + \partial_r)
r^2 (\partial_t + \partial_r)\Psi_2(t,r)
\end{equation}
and the assumptions that (ii) $\Psi_2(0,r) = 0 =
(\partial_t\Psi_2)(0,r)$ in a neighborhood of $r_b$
and (iii) $\Psi_2$ obeys the radial wave equation,
we conclude that $f^{(1)}(-r_b) = 0 = f(-r_b)$.
Therefore, as claimed, the $\ell = 2$ case of 
\eqref{eq:SommerfeldRes} holds exactly subject to
our assumption \eqref{eq:IDassume} on the initial data.
Note that, even if \eqref{eq:IDassume} does not hold, 
the last term in \eqref{eq:full_conv} decays exponentially.

Now let us verify \eqref{eq:full_conv}, assuming only the 
outgoing solution (\ref{eq:ell2mult_td}) without any 
restriction on the initial data. That is, we only assume 
that $\Psi_2 = \Psi_2^{(1)}$, with no contribution from
$\Psi_2^{(-1)}$ [cf.~\eqref{eq:defpsil}].
First consider the quadratic polynomial
\begin{equation}
p(z) \equiv \sqrt{\frac{2z}{\pi}}z^2 e^z K_{5/2}(z) = z^2 + 3z + 3,
\qquad p(z_\pm) = 0,
\end{equation}
where $K_\nu(z)$ is the MacDonald function 
(cf.~footnote \ref{fn:MacDonald}). 
We appeal to the form (\ref{eq:ell2mult_td}) of $\Psi_2$, and via 
integration by parts shift all time derivatives off of $f$ and onto 
the exponentials. For generic $z$ (either $z_+$ or $z_-$) the 
calculation gives
\begin{align}
\begin{split}
  \frac{z}{r^2}\int_0^t\exp\left[\frac{z}{r}(t-t')\right]
& \left[f^{(2)}(t'-r)+\frac{3}{r}f^{(1)}(t'-r)+\frac{3}{r^2}f(t'-r)
  \right]\mathrm{d}t' \\ 
& = 
  \frac{z}{r^2}f^{(1)}(t-r) 
+ \frac{1}{r^3}\big(z^2+3z\big)f(t-r)
\\ 
& - \exp\left(\frac{z}{r}t\right)\left[
  \frac{z}{r^2}f^{(1)}(-r)
 +\frac{1}{r^3}\big(z^2+3z\big)f(-r)\right]
\\
& + (z^2+3z+3)\frac{z}{r^4}\int_0^t \exp\left[\frac{z}{r}(t-t')\right]
f(t'-r)\mathrm{d}t'.
\end{split}
\end{align}
Since $z$ is either $z_+$ or $z_-$, the prefactor in the last 
term is $p(z_\pm) = 0$. Therefore,
\begin{align}
\begin{split}
\frac{z_+}{r^2}\int_0^t
\exp & \left[\frac{z_+}{r}(t-t')\right] \Psi_2(t',r)\mathrm{d}t'
+ \frac{z_-}{r^2}\int_0^t
\exp\left[\frac{z_-}{r}(t-t')\right]\Psi_2(t',r)\mathrm{d}t'
\\
& = \frac{1}{r^2}(z_+ + z_-)f^{(1)}(t-r)
+\frac{1}{r^3}(z_+^2 + 3z_+ +z_-^2 + 3z_-)f(t-r)
\\
&
- \exp\left(\frac{z_+}{r}t\right)\left[
  \frac{z_+}{r^2}f^{(1)}(-r)
 +\frac{1}{r^3}\big(z_+^2+3z_+\big)f(-r)\right]
\\
&
- \exp\left(\frac{z_-}{r}t\right)\left[
  \frac{z_-}{r^2}f^{(1)}(-r)
 +\frac{1}{r^3}\big(z_-^2+3z_-\big)f(-r)\right].
\end{split}
\end{align}
Combination of the last result with 
$z_\pm = -\frac{1}{2}(3 \mp \mathrm{i}\sqrt{3})$ then gives
\begin{align}
& \frac{1}{r}\int_0^t \Omega_2(t-t',r)\Psi_2(t',r)dt'
= -\frac{3}{r^2}f^{(1)}(t-r) -\frac{6}{r^3}f(t-r)
\label{eq:detailsROBCl2}\\
& +\exp\Big(-\frac{3t}{2r}\Big)\Big\{
\sin\Big(\frac{\sqrt{3}t}{2r}\Big)
\Big[\frac{\sqrt{3}}{r^2}f^{(1)}(-r)
\Big]
+ \cos\Big(\frac{\sqrt{3}t}{2r}\Big)
\Big[\frac{3}{r^2}f^{(1)}(-r)
     + \frac{6}{r^3}f(-r)
\Big]\Big\},
\nonumber
\end{align}
from which we immediately get \eqref{eq:full_conv}.

\section{RBC for other foliations of Schwarzschild} 
\label{sec:OtherFol}
In terms of the standard time slices and area radius
the Schwarzschild line-element is
\begin{equation}
ds^2 = -fdt^2 + f^{-1}dr^2 + r^2 d\theta^2  
+ \sin^2\theta d\phi^2,
\qquad f \equiv 1 - 2M/r.
\end{equation}
Define the outgoing (future and outward pointing) null vector 
\begin{equation}
z^+ \equiv \frac{1}{f^{1/2}}\frac{\partial}{\partial t} 
+ f^{1/2}\frac{\partial}{\partial r}
\implies
f^{1/2}z^+ = \frac{\partial}{\partial t} + \frac{\partial}{\partial x},
\end{equation}
again where $x = r_*$ is the tortoise coordinate. The exact RBC for 
these coordinates is essentially \eqref{eq:BH_RBC}, with appropriate 
rescalings by $2M$ factors. In particular, with $\Omega_\ell(t,r) 
= (2M)^{-1}\omega_\ell(t/(2M),r/(2M))$, we write the RBC as
\begin{equation}
f^{1/2}z^+[\Psi] = r^{-1}f(\Omega_\ell * \Psi),
\end{equation}
where $f^{1/2}z^+[\Psi] = X$ from \eqref{eq:charvars}. 
To implement the boundary condition, we approximate it 
through the replacement $\Omega_\ell \rightarrow \Xi_\ell$.

Following Zengino\u{g}lu's \cite{Zenginoglu:2009ey} analysis,
we now consider a change of time slices defined by the new
time variable
\begin{equation}
\lambda = t - h(r),
\end{equation}
where $h(r)$ is the {\em height function}. In terms of $\lambda$ the 
line-element becomes
\begin{equation}
ds^2 = -N^2 d\lambda^2 + g_{rr}
(dr + V^r d\lambda)(dr + V^r d\lambda)
+ r^2 d\theta^2
+ \sin^2\theta d\phi^2,
\end{equation}
where the lapse, radial lapse, and radial component of the
shift vector are respectively
\begin{equation}
N^2 = \frac{f}{1-(fH)^2},\qquad 
\sqrt{g_{rr}} = \frac{1}{N},\qquad 
V^r = -fH N^2.
\end{equation}
Here $H = dh/dr$ is the derivative of the height function.
Define the outgoing ($+$) and incoming ($-$) null vectors
\begin{equation}
w^\pm \equiv \frac{1}{N}\frac{\partial}{\partial \lambda}
-\left(\frac{V^r}{N} \mp \frac{1}{\sqrt{g_{rr}}}\right)
\frac{\partial}{\partial r}.
\end{equation}
Then $z^+ = \exp(\vartheta) w^+$ and $w^+ = \exp(-\vartheta) z^+$, 
where the boost angle is
\begin{equation}
\vartheta = \frac{1}{2}
\log\left[
\frac{1 + \sqrt{g_{rr}} N^{-1}V^r}{1 - \sqrt{g_{rr}}N^{-1} V^r}
\right]
= \frac{1}{2}\log\left[
\frac{1 - N fH \sqrt{1-(fH)^2}}{1 + N fH \sqrt{1-(fH)^2}}
\right].
\end{equation}
Therefore, with respect to the new slices the exact RBC is
\begin{equation}\label{eq:EFconv}
w^+[\Psi] = r^{-1} e^{-\vartheta} f^{1/2}(\Omega_\ell * \Psi),
\end{equation}
and it can similarly be approximated through the replacement
$\Omega_\ell \rightarrow \Xi_\ell$. As given by Zengino\u{g}lu
\cite{Zenginoglu:2009ey}, the $H$ functions for {\em ingoing 
Eddington-Finkelstein} and {\em constant mean curvature} 
foliations are respectively
\begin{equation}
H_\mathrm{iEF} = -\frac{2M}{r-2M},\qquad
H_\mathrm{CMC} = \frac{J}{f\sqrt{J^2 + f}},
\end{equation}
where $J \equiv \frac{1}{3}Kr - Cr^{-2}$ in terms of the
trace $K$ of the extrinsic curvature tensor (based on 
Wald's definition \cite{Wald} of the tensor) and constant 
$C$ of integration.

\section{Numerical Tables} \label{App:Tables}
This appendix collects the tables used for the numerical
simulation documented in Subsection~\ref{subsec:QNRTails}.
Table \ref{tab:appendixRBCtable} determines the kernel 
$\Xi_2(t,30M)$ which approximates the exact kernel
$\Omega_2(t,30M) = (2M)^{-1}\omega_2(t/(2M),15)$. The 19
locations $\beta_{2,q}$ and strengths $\gamma_{2,q}$
which make up this table have 
been computed in quadruple precision and satisfy the 
tolerance \cite{Lau3} $\varepsilon = 10^{-15}$. Entries 
of {\tt 0.00e+00} correspond to outputs from the 
Alpert-Greengard-Hagstrom compression algorithm which are 
typically in the range $10^{-70}$ to $10^{-100}$. 

We provide two different approximations for the time-domain
teleportation kernel $\Phi_2(t,30M,2M(1\times 10^{15})) =
(2M)^{-1}\phi_2(t/(2M),15,1\times 10^{15})$, each denoted
$\Xi^E_2(t,30M,2M(1\times 10^{15}))$. 
Table \ref{tab:appendixEXTtableShort}
determines the first $\Xi^E_2(t,30M,2M(1\times 10^{15}))$. For
this table notice that the 19 locations $\beta^E_{2,q}$ exactly 
match the $\beta_{2,q}$ listed in Table \ref{tab:appendixRBCtable}.
Therefore, with this table the teleportation can be performed
without evolving supplemental convolutions. However, we believe
the tolerance for this table is only 
$\varepsilon = 5 \times 10^{-9}$. Table \ref{tab:appendixEXTtableLong}
determines the second $\Xi^E_2(t,30M,2M(1\times 10^{15}))$ which
now has 26 locations $\beta^E_{2,q}$ and strengths $\gamma^E_{2,q}$. 
Use of this table for teleportation with the RBC specified by 
Table \ref{tab:appendixRBCtable} requires the evolution of 
26 extra convolutions. However, we believe that this second approximate 
kernel satisfies a tolerance of $\varepsilon = 2 \times 10^{-14}$.
\begin{table}
{\scriptsize \begin{Verbatim}[frame=single,framerule=0.2pt,framesep=5pt]
 Regge-Wheeler RBC table for ell = 2 and rho = 15.0 
 Gamma strengths                                  Beta locations
-2.6076002831928367e-08 +0.00e+00                -5.4146529341487581e-01 +0.00e+00 
-1.7937477396220654e-06 +0.00e+00                -4.1310954989396476e-01 +0.00e+00 
-3.2816441859083765e-05 +0.00e+00                -3.1911338482076557e-01 +0.00e+00 
-2.8179763264971427e-04 +0.00e+00                -2.4711219871899659e-01 +0.00e+00 
-1.4509759948015657e-03 +0.00e+00                -1.9108163722923471e-01 +0.00e+00 
-4.4918693070976545e-03 +0.00e+00                -1.4749601558718450e-01 +0.00e+00 
-5.6790046261682662e-03 +0.00e+00                -1.1366299945908588e-01 +0.00e+00 
-2.0012016782502274e-03 +0.00e+00                -8.6476935381164341e-02 +0.00e+00 
-2.9649254206011509e-04 +0.00e+00                -6.4512065175451036e-02 +0.00e+00 
-3.2913867328382246e-05 +0.00e+00                -4.7332374442044557e-02 +0.00e+00 
-3.2675049152330702e-06 +0.00e+00                -3.4115775484663602e-02 +0.00e+00 
-2.8887585153331239e-07 +0.00e+00                -2.4048935704759654e-02 +0.00e+00 
-2.1640495893086479e-08 +0.00e+00                -1.6468632919283480e-02 +0.00e+00 
-1.2772861871474360e-09 +0.00e+00                -1.0845690423058696e-02 +0.00e+00 
-5.3164468909323526e-11 +0.00e+00                -6.7552918597864947e-03 +0.00e+00 
-1.2736896522814067e-12 +0.00e+00                -3.8525630196891325e-03 +0.00e+00 
-1.0598024220301938e-14 +0.00e+00                -1.8481215040788866e-03 +0.00e+00 
-8.9530431033189126e-02 +6.2063746326002998e-02  -9.4779490815239023e-02 +5.9927979877488720e-02 
-8.9530431033189126e-02 -6.2063746326002998e-02  -9.4779490815239023e-02 -5.9927979877488720e-02 
\end{Verbatim}
}
\caption{{\sc Radiation boundary conditions.} As indicated this 
table corresponds the $r_b = 30M$ and $\ell = 2$.
\label{tab:appendixRBCtable}}
\end{table}
\begin{table}
{\scriptsize \begin{Verbatim}[frame=single,framerule=0.2pt,framesep=5pt]
 Regge-Wheeler extraction table for ell = 2 and rho1 = 15.0 to rho2 = 1.0e+15 
 GammaE strengths                                 BetaE locations
-1.7576263057679588e-08 +0.00e+00                -5.4146529341487581e-01 +0.00e+00 
-6.4180514293201244e-08 +0.00e+00                -4.1310954989396476e-01 +0.00e+00 
-6.2732971050093645e-06 +0.00e+00                -3.1911338482076557e-01 +0.00e+00 
-6.9363117988987985e-05 +0.00e+00                -2.4711219871899659e-01 +0.00e+00 
-5.7180637750793345e-04 +0.00e+00                -1.9108163722923471e-01 +0.00e+00 
-2.7884247577175825e-03 +0.00e+00                -1.4749601558718450e-01 +0.00e+00 
-5.8836792033570406e-03 +0.00e+00                -1.1366299945908588e-01 +0.00e+00 
-3.6549136132892194e-03 +0.00e+00                -8.6476935381164341e-02 +0.00e+00 
-1.0498746767499628e-03 +0.00e+00                -6.4512065175451036e-02 +0.00e+00 
-2.4204781878995181e-04 +0.00e+00                -4.7332374442044557e-02 +0.00e+00 
-5.5724464176629910e-05 +0.00e+00                -3.4115775484663602e-02 +0.00e+00 
-1.2157296793548960e-05 +0.00e+00                -2.4048935704759654e-02 +0.00e+00 
-2.6651813247193486e-06 +0.00e+00                -1.6468632919283480e-02 +0.00e+00 
-4.8661708981182769e-07 +0.00e+00                -1.0845690423058696e-02 +0.00e+00 
-8.6183677612060044e-08 +0.00e+00                -6.7552918597864947e-03 +0.00e+00 
-9.3735071189910810e-09 +0.00e+00                -3.8525630196891325e-03 +0.00e+00 
-8.7881787023094076e-10 +0.00e+00                -1.8481215040788866e-03 +0.00e+00 
-9.1164536027591433e-02 -5.3953709155198780e-02  -9.4779490815239023e-02 +5.9927979877488720e-02 
-9.1164536027591433e-02 +5.3953709155198780e-02  -9.4779490815239023e-02 -5.9927979877488720e-02 
\end{Verbatim}
}
\caption{{\sc Teleportation table for AWE.} Note that the locations
in this table match those in Table \ref{tab:appendixRBCtable}.
\label{tab:appendixEXTtableShort}}
\end{table}
\begin{table}
{\scriptsize \begin{Verbatim}[frame=single,framerule=0.2pt,framesep=5pt]
 Regge-Wheeler extraction table for ell = 2 and rho1 = 15.0 to rho2 = 1.0e+15 
 GammaE strengths                                 BetaE locations
-2.7644898994070847e-08 +0.00e+00                -4.7566048766905883e-01 +0.00e+00 
-2.0673535427889927e-06 +0.00e+00                -3.5108913199590891e-01 +0.00e+00 
-4.2379338886728862e-05 +0.00e+00                -2.6424678470854152e-01 +0.00e+00 
-4.3426207863682094e-04 +0.00e+00                -2.0004528999326085e-01 +0.00e+00 
-2.5469795864949394e-03 +0.00e+00                -1.5187870767201261e-01 +0.00e+00 
-6.0248402659888447e-03 +0.00e+00                -1.1566719641292256e-01 +0.00e+00 
-3.8754717050848465e-03 +0.00e+00                -8.7364683652498221e-02 +0.00e+00 
-1.0839648026423402e-03 +0.00e+00                -6.4996045764038071e-02 +0.00e+00 
-2.5061924186508839e-04 +0.00e+00                -4.7886373485938064e-02 +0.00e+00 
-5.8503552767044972e-05 +0.00e+00                -3.4988356300604928e-02 +0.00e+00 
-1.4014861186166239e-05 +0.00e+00                -2.5326730622776284e-02 +0.00e+00 
-3.3862132702752687e-06 +0.00e+00                -1.8135330085424492e-02 +0.00e+00 
-8.0926034459750484e-07 +0.00e+00                -1.2824248788413890e-02 +0.00e+00 
-1.8800680938314195e-07 +0.00e+00                -8.9386729575681410e-03 +0.00e+00 
-4.1796375696032426e-08 +0.00e+00                -6.1274292303408907e-03 +0.00e+00 
-8.7558713932917318e-09 +0.00e+00                -4.1196547786942856e-03 +0.00e+00 
-1.7001775199292710e-09 +0.00e+00                -2.7071621335894381e-03 +0.00e+00 
-3.0014940247349529e-10 +0.00e+00                -1.7308275437738592e-03 +0.00e+00 
-4.7007801744202052e-11 +0.00e+00                -1.0699222812596474e-03 +0.00e+00 
-6.3144518165051684e-12 +0.00e+00                -6.3367274917212608e-04 +0.00e+00 
-6.9169662522379380e-13 +0.00e+00                -3.5455566302145149e-04 +0.00e+00 
-5.6820400420323507e-14 +0.00e+00                -1.8296880288072223e-04 +0.00e+00 
-2.9748099840520155e-15 +0.00e+00                -8.2999833632165545e-05 +0.00e+00 
-6.5083672189429277e-17 +0.00e+00                -2.9042514390090429e-05 +0.00e+00 
-9.1164550073798437e-02 -5.3953205902806563e-02  -9.4779494659287755e-02 +5.9928005360963245e-02 
-9.1164550073798437e-02 +5.3953205902806563e-02  -9.4779494659287755e-02 -5.9928005360963245e-02 
\end{Verbatim}
}
\caption{{\sc Teleportation table for AWE.} This table
is a more accurate approximation to the kernel also approximated
by Table \ref{tab:appendixEXTtableShort}.
\label{tab:appendixEXTtableLong}}
\end{table}

\end{document}